%% file: InfoDynamicsToolkit.tex
\pdfoutput=1

\newif\ifarXiv
\arXivtrue 

\ifarXiv

\documentclass[
 aps, 
 amsmath,amssymb,
 pre,superscriptaddress,showpacs, 
 showkeys, 
 numerical,
 nofootinbib
]{revtex4-1}

\else

\documentclass{frontiersENG} 

\usepackage{url,lineno}
\linenumbers

\copyrightyear{}
\pubyear{}

\def\keyFont{\fontsize{8}{11}\helveticabold }
\def\firstAuthorLast{Lizier} 

\def\Authors{Joseph T. Lizier\,$^{1,2,*}$}

\fi

\newcommand{\fig}[1]{Fig.~\ref{fig:#1}}
\newcommand{\eq}[1]{Eq.~(\ref{eq:#1})}
\newcommand{\eqs}[2]{Eq.~(\ref{eq:#1},\ref{eq:#2})}

\newcommand{\secRef}[1]{Section \ref{sec:#1}}

\ifarXiv
\newcommand{\appRef}{Appendix}
\newcommand{\app}[1]{Appendix \ref{app:#1}}
\else
\newcommand{\appRef}{Supplementary Material}
\newcommand{\app}[1]{Supplementary Material \S\ref{app:#1}}
\fi
\newcommand{\appRefAbbrev}[1]{\S\ref{app:#1}}
\newcommand{\appRefSec}[1]{Section \ref{app:#1}}
\newcommand{\tableRef}[1]{Table \ref{table:#1}}
\newcommand{\listRef}[1]{Listing \ref{list:#1}}
\newcommand{\lineRef}[1]{line \ref{line:#1}}

\newcommand{\etal}{\textsl{et al.}}
\newcommand{\caWidth}{0.23\textwidth}
\newcommand{\caHeight}{0.17\textwidth}

\newcommand{\entropy}[1]{H(#1)}
\newcommand{\jointEntropy}[2]{H(#1,#2)}
\newcommand{\condEntropy}[2]{H(#1 \mid #2)}
\newcommand{\mutualInfo}[2]{I(#1 ; #2)}
\newcommand{\mutualInfoUpper}[3]{I^{#1}(#2 ; #3)}
\newcommand{\multiInfo}[1]{I(#1)}
\newcommand{\multiInfoPlus}[3]{I(#1 ; #2 ; \ldots ; #3)}
\newcommand{\condMutualInfo}[3]{I(#1 ; #2 \mid #3)}
\newcommand{\conditionalMutualInfoUpper}[4]{I^{#1}(#2 ; #3 \mid #4)}
\newcommand{\entropyRate}[1]{H'_{\mu #1}}

\newcommand{\entropyRateCond}[1]{H_{\mu #1}}
\newcommand{\entropyRateCondK}[2]{H_{\mu #1}(#2)}
\newcommand{\excessEntropy}[1]{E_{#1}}
\newcommand{\excessEntropyK}[2]{E_{#1}(#2)}
\newcommand{\ais}[1]{A_{#1}}
\newcommand{\aisK}[2]{A_{#1}(#2)}
\newcommand{\te}[2]{T_{#1 \rightarrow #2}}
\newcommand{\teArgs}[3]{T_{#1 \rightarrow #2}(#3)}

\newcommand{\condTe}[3]{T_{#1 \rightarrow #2 \mid #3}}
\newcommand{\condTeArgs}[4]{T_{#1 \rightarrow #2 \mid #3}(#4)}
\newcommand{\condTeK}[4]{T_{#1 \rightarrow #2 \mid #3}(#4)}

\newcommand{\separable}[1]{S_{#1}}
\newcommand{\separableK}[2]{S_{#1}(#2)}
\newcommand{\localEntropy}[1]{h(#1)}
\newcommand{\localJointEntropy}[2]{h(#1,#2)}
\newcommand{\localCondEntropy}[2]{h(#1 \mid #2)}
\newcommand{\localMutualInfo}[2]{i(#1 ; #2)}
\newcommand{\localMutualInfoUpper}[3]{i^{#1}(#2 ; #3)}

\newcommand{\localMultiInfoPlus}[3]{i(#1 ; #2 ; \ldots ; #3)}
\newcommand{\localCondMutualInfo}[3]{i(#1 ; #2 \mid #3)}
\newcommand{\localCondMutualInfoUpper}[4]{i^{#1}(#2 ; #3 \mid #4)}
\newcommand{\localEntropyRateCond}[2]{h_{\mu #1}(#2)}
\newcommand{\localExcessEntropy}[2]{e_{#1}(#2)}
\newcommand{\localAis}[2]{a_{#1}(#2)}
\newcommand{\localTe}[3]{t_{#1 \rightarrow #2}(#3)}
\newcommand{\localCondTe}[4]{t_{#1 \rightarrow #2 \mid #3}(#4)}
\newcommand{\diffEntropy}[1]{H_D(#1)}
\newcommand{\condDiffEntropy}[2]{H_D(#1 \mid #2)}
\newcommand{\diffMI}[2]{I_D(#1 ; #2)}

\usepackage{graphicx}
\usepackage{subfigure} 
\usepackage{supertabular,array}
\usepackage{hyphenat} 
\usepackage{multirow}
\usepackage[hide]{todo}
\usepackage{amssymb}  
\usepackage{listings} 
\usepackage{xcolor} 
\colorlet{light-gray}{gray!20}

\ifarXiv
\else
\usepackage{xr}
\fi

\newcommand{\paperTitle}{JIDT: An information-theoretic toolkit for studying the dynamics of complex systems}
\newcommand{\theKeywords}{information transfer, information storage, intrinsic computation, complex networks, information theory, transfer entropy, Java, MATLAB, Octave, Python}

\ifx\pdfoutput\undefined
\else
    \usepackage[bookmarks=true,colorlinks=false,citecolor=black,linkcolor=black,breaklinks=true,pdftex]{hyperref}
  	\hypersetup{
      	pdfauthor = {Joseph T. Lizier},
        pdftitle = {\paperTitle},
  	    pdfsubject = {Version 3.1; svn 267},
      	pdfkeywords = {\theKeywords},
        pdfcreator = {LaTeX with hyperref package},
  	    pdfproducer = {pdftex},
  	    pdfstartview = {FitH}}
\fi

\begin{document}

\ifarXiv

\title{\paperTitle}

\author{Joseph T. Lizier}
\email[]{joseph.lizier@gmail.com}
\thanks{This manuscript is a modified pre-print of: Joseph T. Lizier, ``JIDT: An information-theoretic toolkit for studying the dynamics of complex systems'', \textit{Frontiers in Robotics and AI} 1:11, 2014; doi:\href{http://dx.doi.org/10.3389/frobt.2014.00011}{10.3389/frobt.2014.00011}}
\affiliation{Max Planck Institute for Mathematics in the Sciences,
Inselstra{\ss}e 22, 04103 Leipzig, Germany}
\affiliation{CSIRO Digital Productivity Flagship, Marsfield, NSW 2122, Australia}

\date{\today}

\else

\onecolumn
\firstpage{1}

\title[\paperTitle]{\paperTitle} 
\author[\firstAuthorLast ]{\Authors}
\address{}
\correspondance{}
\extraAuth{}
\topic{}

\maketitle

\fi

\begin{abstract}
Complex systems are increasingly being viewed as distributed information processing systems, particularly in the domains of computational neuroscience, bioinformatics and Artificial Life.
This trend has resulted in a strong uptake in the use of (Shannon) information-theoretic measures to analyse the dynamics of complex systems in these fields.
We introduce the Java Information Dynamics Toolkit (JIDT): a Google code project which provides a standalone, (GNU GPL v3 licensed) open-source code implementation for empirical estimation of information-theoretic measures from time-series data.
While the toolkit provides classic information-theoretic measures (e.g. entropy, mutual information, conditional mutual information), it ultimately focusses on implementing higher-level measures for information dynamics.
That is, JIDT focusses on quantifying information storage, transfer and modification,  and the dynamics of these operations in space and time.
For this purpose, it includes implementations of the transfer entropy and active information storage, their multivariate extensions and local or pointwise variants.
JIDT provides implementations for both discrete and continuous-valued data for each measure, including various types of estimator for continuous data (e.g. Gaussian, box-kernel and Kraskov-St\"ogbauer-Grassberger) which can be swapped at run-time due to Java's object-oriented polymorphism.
Furthermore, while written in Java, the toolkit can be used directly in MATLAB, GNU Octave, Python and other environments.
We present the principles behind the code design, and provide several examples to guide users.
\ifarXiv\else
\tiny
 \keyFont{ \section{Keywords:} \theKeywords} 
\fi
\end{abstract}

\ifarXiv

\maketitle 

\fi

\section{Introduction}
\label{sec:intro}

\emph{Information theory} was originally introduced by \citet{shan48} to quantify fundamental limits on signal processing operations and reliable communication of data \citep{cover91,mac03}.
More recently, it is increasingly being utilised for the design and analysis of complex self-organized systems \citep{pro09}.
\emph{Complex systems} science \citep{mitch09} is the study of large collections of entities (of some type), where the global system behaviour is a non-trivial result of the local interactions of the individuals; e.g. emergent consciousness from neurons, emergent cell behaviour from gene regulatory networks and flocks determining their collective heading.
The application of information theory to complex systems can be traced to the increasingly-popular perspective that commonalities between complex systems may be found ``in the way they handle information'' \citep{gell94}.
Certainly, there have been many interesting insights gained from the application of traditional information-theoretic measures such as entropy and mutual information to study complex systems, for example: proposals of candidate complexity measures \citep{ton94,adami02}, characterising order-chaos phase transitions \citep{pro05a,mira95,sole01,pro11a}, and measures of network structure \citep{sole04,pir09b}.

More specifically though, researchers are increasingly viewing the global behaviour of complex systems as emerging from the \emph{distributed information processing}, or \emph{distributed computation}, between the individual elements of the system \citep{liz13a,lang90,fern06a,mitch98a,wang12a}, e.g. collective information processing by neurons \citep{gong09}.
Computation in complex systems is examined in terms of:
how information is transferred in the interaction between elements, how it is stored by elements, and how these information sources are non-trivially combined.
We refer to the study of these operations of \emph{information storage, transfer and modification}, and in particular how they unfold in space and time, as \emph{information dynamics} \citep{liz13a,liz14a}.

Information theory is the natural domain to quantify these operations of information processing, and we have seen a number of measures recently introduced for this purpose, including the well-known transfer entropy \citep{schr00}, as well as active information storage \citep{liz12a} and predictive information \citep{bialek01,crutch03}.
Natural affinity aside, information theory offers several distinct advantages as a measure of information processing in dynamics,\footnote{Further commentary on links between information-theoretic analysis and traditional dynamical systems approaches are discussed by \citet{beer14a}.} including: its model-free nature (requiring only access to probability distributions of the dynamics), ability to handle stochastic dynamics and capture non-linear relationships, its abstract nature, generality, and mathematical soundness.

In particular, this type of information-theoretic analysis has gained a strong following in computational neuroscience, where the transfer entropy has been widely applied \citep{hon07,wib14b,wib14c,mar12b,stra12a,faes11a,liz11a,vic11a,stett12a,vak09,ito11a,maki13,liao11},  (for example for effective network inference), and measures of information storage are gaining traction \citep{wib14a,faes14a,gomez14a}.
Similarly, such information-theoretic analysis is popular in studies of canonical complex systems \citep{mahon11a,barn13a,wang12a,liz08a,liz10e,liz12a}, dynamics of complex networks \citep{liz08c,dam10a,dam11a,san14a,liz11b,liz12b}, social media \citep{steeg13a,oka13a,bauer12b}, and in Artificial Life and Modular Robotics both for analysis \citep{pro06a,lung06,nak12a,nak13a,will10b,obst13a,boed12a,cliff13a,walk12a,liz09d} and design \citep{pro06b,liz08d,kly08a,ay08a,das13a,obs10a} of embodied cognitive systems (in particular see the ``Guided Self-Organization'' series of workshops, e.g. \citep{pro09b}).

This paper introduces \textbf{JIDT} -- the \textbf{Java Information Dynamics Toolkit} -- which provides a standalone implementation of information-theoretic measures of dynamics of complex systems.
JIDT is open-source, licensed under GNU General Public License v3, and available for download via Google code at \href{http://code.google.com/p/information-dynamics-toolkit/}{http://code.google.com/p/information-dynamics-toolkit/}.
JIDT is designed to facilitate \emph{general purpose} empirical estimation of information-theoretic measures from time-series data, by providing easy to use, portable implementations of measures of information transfer, storage, shared information and entropy.

We begin by describing the various information-theoretic measures which are implemented in JIDT in \secRef{measures} and \app{measures}, including the basic entropy and (conditional) mutual information \citep{cover91,mac03}, as well as the active information storage \citep{liz12a}, the transfer entropy \citep{schr00} and its conditional/multivariate forms \citep{liz08a,liz10e}.
We also describe how one can compute \emph{local} or \emph{pointwise} values of these information-theoretic measures at specific observations of time-series processes, so as to construct their \emph{dynamics} in time.
We continue to then describe the various estimator types which are implemented for each of these measures in \secRef{estimationTechniques} and \app{estimationTechniques} (i.e. for discrete or binned data, and Gaussian, box-kernel and Kraskov-St\"ogbauer-Grassberger estimators).
Readers familiar with these measures and their estimation may wish to skip these sections.
We also summarise the capabilities of similar information-theoretic toolkits in \secRef{otherToolkits} (focussing on those implementing the transfer entropy).

We then turn our attention to providing a detailed introduction of JIDT in \secRef{toolkit}, focussing on the current version 1.0 distribution.
We begin by highlighting the unique features of JIDT in comparison to related toolkits, in particular in: providing \textit{local} information-theoretic measurements of dynamics; implementing conditional and other multivariate transfer entropy measures; and including implementations of other related measures including the active information storage.
We describe the (almost zero) installation process for JIDT in \secRef{installation}:
JIDT is standalone software, requiring no prior installation of other software (except a Java Virtual Machine), and no explicit compiling or building.
We describe the contents of the JIDT distribution in \secRef{contents}, and then in \secRef{measuresImplemented} outline which estimators are implemented for each information-theoretic measure.
We then describe the principles behind the design of the toolkit in \secRef{architecture}, including our object-oriented approach in defining interfaces for each measure, then providing multiple implementations (one for each estimator type).
\secRef{validation}-\secRef{documentation} then describe how the code has been tested, how the user can (re-)build it, and what extra documentation is available (principally the project wiki and Javadocs).

Finally and most importantly, \secRef{demos} outlines several demonstrative examples supplied with the toolkit, which are intended to guide the user through how to use JIDT in their code.
We begin with simple Java examples in \secRef{simpleJavaDemos}, which includes a description of the general pattern of usage in instantiating a measure and making calculations, and walks the user through differences in calculators for discrete and continuous data, and multivariate calculations.
We also describe how to take advantage of the polymorphism in JIDT's object-oriented design to facilitate run-time swapping of the estimator type for a given measure.
Other demonstration sets from the distribution are presented also, including: basic examples using the toolkit in MATLAB, GNU Octave and Python (\secRef{matlabDemos} and \secRef{pythonDemos}); reproduction of the original transfer entropy examples from \citet{schr00} (\secRef{schreiberDemos}); and local information profiles for cellular automata (\secRef{caDemos}).

\section{Information-theoretic measures and estimators}

We begin by providing brief overviews of information-theoretic measures (\secRef{measures}) and estimator types (\secRef{estimationTechniques}) implemented in JIDT. These sections serve as summaries of \app{measures} and \app{estimationTechniques}.
We also discuss related toolkits implementing some of these measures in \secRef{otherToolkits}.

\subsection{Information-theoretic measures}
\label{sec:measures}

This section provides a brief overview of the information-theoretic measures  \citep{cover91,mac03} which are implemented in JIDT. All features discussed are available in JIDT unless otherwise noted.
\emph{A more complete description for each measure is provided in} \app{measures}.

We consider measurements $x$ of a random variable $X$, with a probability distribution function (PDF) $p(x)$ defined over the alphabet $\alpha_x$ of possible outcomes for $x$ (where $\alpha_x = \{0,\ldots,M_X-1\}$ without loss of generality for some $M_X$ discrete symbols).

The fundamental quantity of information theory for example is the \textbf{Shannon entropy}, which represents the expected or average uncertainty associated with any measurement $x$ of $X$:
\begin{align}
	\entropy{X} = -\sum_{x \in \alpha_x} p(x) \log_2{p(x)}
	\label{eq:entropyMain}.
\end{align}
Unless otherwise stated, logarithms are taken by convention in base 2, giving units in bits.
$\entropy{X}$ for a measurement $x$ of $X$ can also be interpreted as the minimal expected or average number of bits required to encode or describe its value without losing information \citep{mac03,cover91}.
$X$ may be a joint or vector variable, e.g. $\mathbf{X} = \{Y,Z\}$, generalising \eq{entropyMain} to the \textbf{joint entropy} $\entropy{\mathbf{X}}$ or $\jointEntropy{Y}{Z}$ for an arbitrary number of joint variables (see \tableRef{measuresSummary} and \eq{jointEntropy} in \app{classicMeasures}).
While the above definition of Shannon entropy applies to discrete variables, it may be extended to variables in the \textit{continuous} domain as the \textbf{differential entropy} -- see \app{differentialEntropy} for details.

All of the subsequent Shannon information-theoretic quantities we consider may be written as sums and differences of the aforementioned marginal and joint entropies, and all may be extended to multivariate ($\mathbf{X}$, $\mathbf{Y}$ etc) and/or continuous variables.
The basic information-theoretic quantities: \textbf{entropy}, \textbf{joint entropy}, \textbf{conditional entropy}, \textbf{mutual information} (MI), \textbf{conditional mutual information} \citep{cover91,mac03}, and \textbf{multi-information} \citep{ton94}; are discussed in detail in \app{classicMeasures}, and summarised here in \tableRef{measuresSummary}.
All of these measures are non-negative.

\begin{table*}[!t]
\caption{Basic information-theoretic quantities (first six rows) and measures of information dynamics (last five rows) implemented in JIDT. Equations are supplied for both their average or expected form, and their local form. References are given to the presentation of these equations in \app{classicMeasures} and \app{infoDynamicsMeasures}.}
\label{table:measuresSummary}
\centering
\ifarXiv\else\begin{footnotesize}\fi
\begin{tabular}{|c||c|c||c|c|}
 \hline
 \textbf{Measure} & \multicolumn{2}{c||}{\textbf{Average/Expected form}} & \multicolumn{2}{c|}{\textbf{Local form}} \\
 \hline
 \hline
 Entropy & $ \entropy{X} = -\sum\limits_{x \in \alpha_x} p(x) \log_2{p(x)} $ & \eq{entropy} & $\localEntropy{x} = - \log_2{p(x)}$ & \eq{localEntropy} \\
 \hline
 Joint entropy & $\jointEntropy{X}{Y} = -\sum\limits_{x \in \alpha_x,y \in \alpha_y}{ p(x,y) \log_2{p(x,y)}}$ & \eq{jointEntropy} & $\localJointEntropy{x}{y} = - \log_2{p(x,y)}$ & \eq{localJointEntropy} \\
 \hline
 Conditional entropy & $\condEntropy{Y}{X} = \jointEntropy{X}{Y} - \entropy{X}$ & \eq{conditionalEntropy} & $\localCondEntropy{x}{y} = \localJointEntropy{x}{y} - \localEntropy{x}$ & \eq{localConditional} \\
 \hline
 Mutual information & $\mutualInfo{X}{Y} = \entropy{X} + \entropy{Y} -\jointEntropy{X}{Y}$ & \eq{mi} & $\localMutualInfo{x}{y} = \localEntropy{x} + \localEntropy{y} -\localJointEntropy{x}{y}$ & \eq{localMi} \\
 \hline
 \multirow{2}{*}{Multi-information} & $\multiInfoPlus{X_1}{X_2}{X_G} = \left( \sum_{g=1}^{G}{\entropy{X_g}} \right)$ \ \ \ \ \ \ \ & \multirow{2}{*}{\eq{multiInfo}} & $\localMultiInfoPlus{x_1}{x_2}{x_G} = \left( \sum_{g=1}^{G}{\localEntropy{x_g}} \right)$ & \multirow{2}{*}{\eq{localMultiInfo}} \\
 & \ \ \ \ \ \ \ \ \ \ \ \ \ \ \ \ \ \ \ \ \ \ \ $ - \entropy{X_1,X_2,\ldots,X_G}$ & & \ \ \ \ \ \ \ \ \ \ \ \ \ \ \ \ \ \ \ \ \ \ \ \ \ \ \ \ \ $ - \localEntropy{x_1,x_2,\ldots,x_G}$ & \\
 \hline
 Conditional MI & $\condMutualInfo{X}{Y}{Z} = \condEntropy{X}{Z} + \condEntropy{Y}{Z}$ & \eq{condMi} & $\localCondMutualInfo{x}{y}{z} = \localCondEntropy{x}{z} + \localCondEntropy{y}{z} $ & \eq{localCondMi} \\
 & \ \ \ \ \ \ \ \ \ \ \ \ $  -\condEntropy{X,Y}{Z}$ & & \ \ \ \ \ \ \ \ \ \ \ \ $ -\localCondEntropy{x,y}{z}$ & \\
 \hline
 \hline
 Entropy rate & $\entropyRateCondK{X}{k} = \condEntropy{X_{n+1}}{\mathbf{X}_{n}^{(k)}}
$ & \eq{entropyRateConditionalKSTat} & $\localEntropyRateCond{X}{n+1,k} = \localCondEntropy{x_{n+1}}{\mathbf{x}_{n}^{(k)}}$ & \eq{localEntropyRateConditionalK} \\
 \hline
 Active information & \multirow{2}{*}{$\aisK{X}{k} = \mutualInfo{\mathbf{X}_n^{(k)}}{X_{n+1}}$} & \multirow{2}{*}{\eq{activeStorageEstimate}} & \multirow{2}{*}{$\localAis{X}{n+1,k} = \localMutualInfo{\mathbf{x}^{(k)}_n}{x_{n+1}}$} & \multirow{2}{*}{\eq{localActiveK}} \\
 storage & & & & \\
 \hline
 Predictive information & $\excessEntropyK{X}{k} = \mutualInfo{\mathbf{X}_n^{(k)}}{\mathbf{X}_{n+1}^{(k^+)}}$ & \eq{predictiveInfoFiniteK} & $\localExcessEntropy{X}{n+1,k} = \localMutualInfo{\mathbf{x}_n^{(k)}}{\mathbf{x}_{n+1}^{(k^+)}}$ & \eq{localPredictiveInfoFiniteK} \\
 \hline
 Transfer entropy & $\teArgs{Y}{X}{k,l,u} = \condMutualInfo{\mathbf{Y}_{n+1-u}^{(l)}}{X_{n+1}}{\mathbf{X}_n^{(k)}}$ & \eq{teDelay} & $\localTe{Y}{X}{n+1, k, l,u} = \localCondMutualInfo{\mathbf{y}^{(l)}_{n+1-u}}{x_{n+1}}{\mathbf{x}^{(k)}_{n}}$ & \eq{localTEK} \\
 \hline
 Conditional TE & $\condTeArgs{Y}{X}{Z}{k,l} = \condMutualInfo{\mathbf{Y}_n^{(l)}}{X_{n+1}}{\mathbf{X}_n^{(k)},Z_n}$ & \eq{teCondEstimate} & $\localCondTe{Y}{X}{Z}{n+1, k, l} = \localCondMutualInfo{\mathbf{y}^{(l)}_{n}}{x_{n+1}}{\mathbf{x}^{(k)}_{n}, z_n}$ & \eq{localCondTEK} \\
 \hline
\end{tabular}
\ifarXiv\else\end{footnotesize}\fi
\end{table*}

Also, we may write down \textbf{pointwise} or \textbf{local information-theoretic measures}, which characterise the information attributed with \emph{specific} measurements $x$, $y$ and $z$ of variables $X$, $Y$ and $Z$ \citep{liz14b}, rather than the traditional expected or average information measures associated with these variables introduced above.
Full details are provided in \app{localMeasures}, and the local form for all of our basic measures is shown here in \tableRef{measuresSummary}.
For example, the \textbf{Shannon information content} or \textbf{local entropy} of an outcome $x$ of measurement of the variable $X$ is \citep{mac03,ash65}: 
\begin{align}
	\localEntropy{x} = - \log_2{p(x)}
	\label{eq:localEntropyMain}.
\end{align}
By convention we use lower-case symbols to denote local information-theoretic measures.
The Shannon information content of a given symbol $x$ is the \emph{code-length} for that symbol in an optimal encoding scheme for the measurements $X$, i.e. one that produces the minimal expected code length.
We can form all local information-theoretic measures as sums and differences of local entropies (see \tableRef{measuresSummary} and \app{localMeasures}), and each ordinary measure is the \emph{average} or \emph{expectation value} of their corresponding local measure, e.g. $\entropy{X} = \left\langle h(x) \right\rangle$.
Crucially, the \textbf{local MI} and \textbf{local conditional MI} \citep[ch. 2]{fano61} may be negative, unlike their averaged forms. This occurs for MI where the measurement of one variable is \textit{misinformative} about the other variable (see further discussion in \app{localMeasures}).

Applied to \textbf{time-series data}, these local variants return a time-series for the given information-theoretic measure, which with mutual information for example characterises how the shared information between the variables fluctuates \textit{as a function of time}.
As such, they directly reveal the \emph{dynamics} of information, and are gaining popularity in complex systems analysis \citep{sha01a,sha06,hel04,wib14a,liz07b,liz08a,liz12a,liz10e,liz14b}.

Continuing with time-series, we then turn our attention to measures specifically used to quantify the dynamics of information processing in multivariate time-series, under a framework for \emph{information dynamics} which was recently introduced by \citet{liz07b,liz08a,liz10e,liz12a,liz14a} and \citet{liz13a,liz14b}.
The measures of information dynamics implemented in JIDT -- which are the real focus of the toolkit -- are discussed in detail in \app{infoDynamicsMeasures}, and summarised here in \tableRef{measuresSummary}.

These measures consider \emph{time-series processes} $X$ of the random variables $\{ \ldots X_{n-1}, X_{n}, X_{n+1} \ldots \}$ with process realisations $\{ \ldots x_{n-1}, x_{n}, x_{n+1} \ldots \}$ for countable time indices $n$.
We use $\mathbf{X}^{(k)}_n = \left\{ X_{n-k+1}, \ldots , X_{n-1}, X_n \right\}$ to denote the $k$ consecutive variables of $X$ up to and including time step \textit{n}, which has realizations $\mathbf{x}^{(k)}_n = \left\{ x_{n-k+1}, \ldots , x_{n-1}, x_n \right\}$.\footnote{We use the corresponding notation $\mathbf{X}_{n+1}^{(k^+)}$ for the next $k$ values \textit{from} $n+1$ onwards, $\left\{ X_{n+1}, X_{n+2}, \ldots , X_{n+k} \right\}$, with realizations $\mathbf{x}_{n+1}^{(k^+)} = \left\{ x_{n+1}, x_{n+2}, \ldots , x_{n+k} \right\}$} The $\mathbf{x}_n^{(k)}$ are Takens' \textit{embedding vectors} \citep{takens81} with \textit{embedding dimension} $k$, which capture the underlying \textit{state} of the process $X$ for Markov processes of order $k$.\footnote{We can use an embedding delay $\tau$ to give $\mathbf{x}_n^{(k)} = \left\{ x_{n-(k-1)\tau}, \ldots , x_{n-\tau}, x_n \right\}$, where this helps to better empirically capture the state from a finite sample size. Non-uniform embeddings (i.e. with irregular delays) may also be useful \citep{faes11a} (not implemented in JIDT at this stage).}

\begin{figure}[t]
	\begin{center}
		\includegraphics[trim= 120 100 50 110,clip=true,width=0.99\columnwidth]{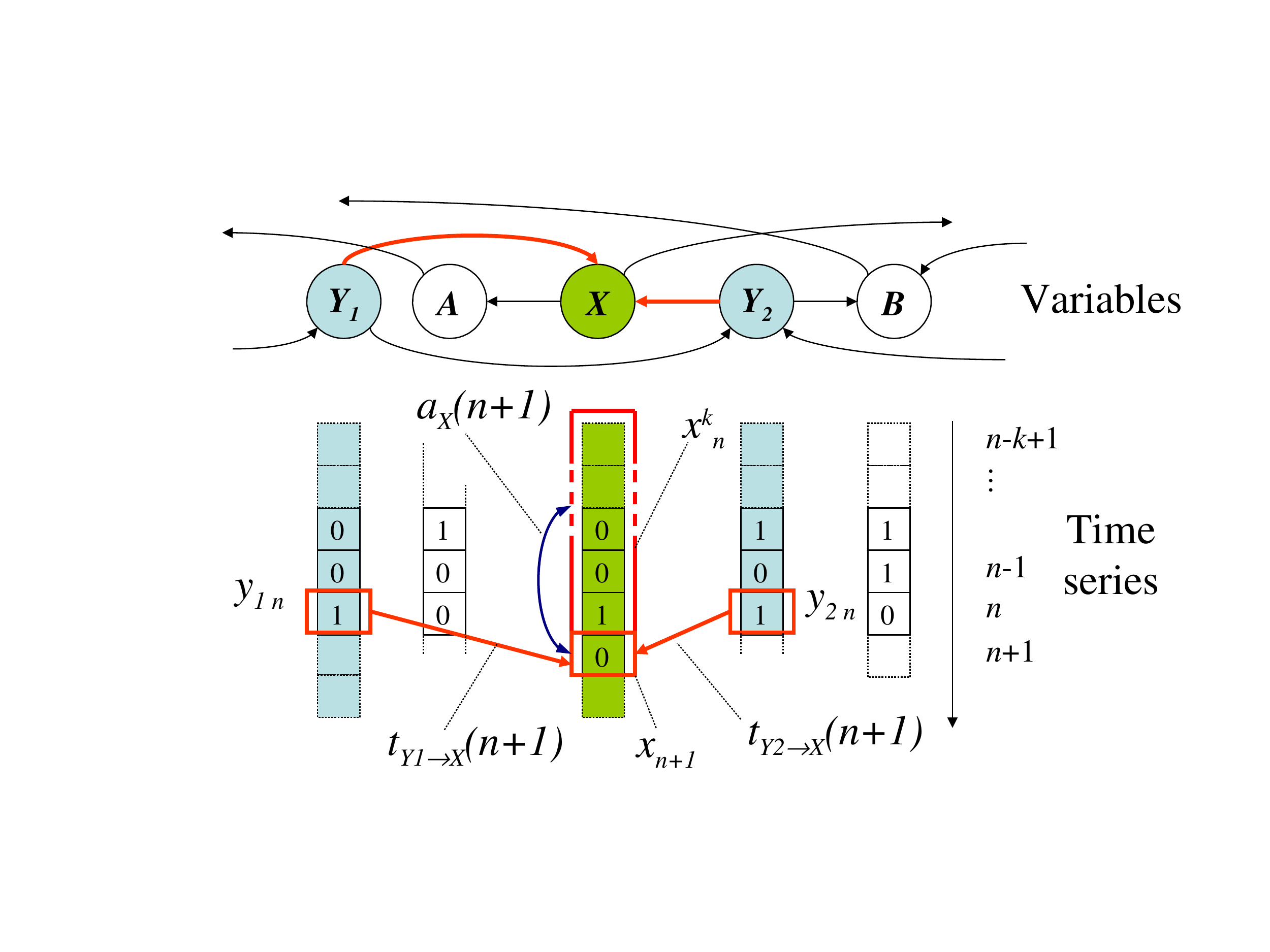}
	\end{center}
	\caption{\label{fig:infoDynamics}
		  Measures of information dynamics with respect to a destination variable $X$. We address the information content in a measurement $x_{n+1}$ of $X$ at time $n+1$ with respect to the active information storage $\localAis{X}{n+1,k}$, and local transfer entropies $\localTe{Y_1}{X}{n+1,k}$ and $\localTe{Y_2}{X}{n+1,k}$ from variables $Y_1$ and $Y_2$.
}
\end{figure}

Specifically, our framework examines how the information in variable $X_{n+1}$ is related to previous variables or states (e.g. $X_{n}$ or $\mathbf{X}^{(k)}_n$) of the process or other related processes, addressing the fundamental question: \emph{``where does the information in a random variable $X_{n+1}$ in a time series come from?''}.
As indicated in \fig{infoDynamics} and shown for the respective measures in \tableRef{measuresSummary}, this question is addressed in terms of:
\begin{enumerate}
	\item information from the past of process $X$ -- i.e. the information \emph{storage}, measured by the \textbf{active information storage} \citep{liz12a} and \textbf{predictive information} or \textbf{excess entropy} \citep{bialek01,crutch03,grass86a};
	\item information contributed from other source processes $Y$ -- i.e. the information \emph{transfer}, measured by the \textbf{transfer entropy} (TE) \citep{schr00} and \textbf{conditional transfer entropy} \citep{liz08a,liz10e};
	\item and how these sources combine -- i.e. information \emph{modification} (see \textbf{separable information} \citep{liz10e} in \app{infoDynamicsMeasures}).
\end{enumerate}
The goal of the framework is to decompose the information in the next observation $X_{n+1}$ of process $X$ in terms of these information sources.

The transfer entropy, arguably the most important measure in the toolkit, has become a very popular tool in complex systems in general, e.g. \citep{will11a,lung06,obs10a,barn12a,liz08a,liz11b,boed12a}, and in computational neuroscience in particular, e.g. \citep{vic11a,lind11a,ito11a,stra12a,liz11a}.
For multivariate Gaussians, the TE is equivalent (up to a factor of 2) to the \textbf{Granger causality} \citep{barn09}.
Extension of the TE to arbitrary source-destination lags is described by \citet{wib13a} and incorporated in \tableRef{measuresSummary} (this is not shown for conditional TE here for simplicity, but is handled in JIDT).
Further, one can consider multivariate sources $\mathbf{Y}$, in which case we refer to the measure $\teArgs{\mathbf{Y}}{X}{k,l}$ as a \textbf{collective transfer entropy} \citep{liz10e}.
See further description of this measure at \app{infoDynamicsMeasures}, including regarding how to set the history length $k$.

\tableRef{measuresSummary} also shows the local variants of each of the above measures of information dynamics (presented in full in \app{localMeasures}).
The use of these local variants is particularly important here because they provide a direct, model-free mechanism to analyse the \emph{dynamics} of how information processing unfolds in time in complex systems.
\fig{infoDynamics} indicates for example a local active information storage measurement for time-series process $X$, and a local transfer entropy measurement from process $Y$ to $X$.

Finally, in \app{statSigTesting} we describe how one can evaluate whether an MI, conditional MI or TE is statistically different from zero, and therefore represents sufficient evidence for a (directed) relationship between the variables.
This is done (following \citep{cha03,verd05,vic11a,lind11a,liz11a,wib14c,barn12a}) via permutation testing to construct appropriate surrogate populations of time-series and measurements under the null hypothesis of no directed relationship between the given variables.

\subsection{Estimation techniques}
\label{sec:estimationTechniques}

While the mathematical formulation of the quantities in \secRef{measures} are relatively straightforward, empirically estimating them in practice from a finite number $N$ of samples of time-series data can be a complex process, and is dependent on the type of data you have and its properties.
Estimators are typically subject to bias and variance due to finite sample size.
Here we briefly introduce the various types of estimators which are included in JIDT, referring the reader to \app{estimationTechniques} (and also \citet{vic14a} for the transfer entropy in particular) for more detailed discussion.

For \textbf{discrete} variables $X$, $Y$, $Z$ etc., the definitions in \secRef{measures} may be used directly by counting the matching configurations in the available data to obtain the relevant plug-in probability estimates (e.g. $\hat{p}(x \mid y)$ and $\hat{p}(x)$ for MI).
These estimators are simple and fast, being implemented in $\operatorname{O}\left(N \right)$ time.
Several bias correction techniques are available, e.g. \citep{pan03,bon08a}, though not yet implemented in JIDT.

For \textbf{continuous} variables $X$, $Y$, $Z$, one could simply discretise or bin the data and apply the discrete estimators above.
While this is simple and fast ($\operatorname{O}\left(N \right)$ as above), it is likely to sacrifice accuracy.
Alternatively, we can use an estimator that harnesses the continuous nature of the variables, dealing with the differential entropy and probability density functions.
The latter is more complicated but yields a more accurate result.
We discuss several such estimators in \app{continuousEstimators}, and summarise them in the following:\footnote{Except where otherwise noted, JIDT implements the most efficient described algorithm for each estimator.}
\begin{itemize}
	\item A \textbf{multivariate Gaussian model} may be used (\app{gaussianEstimators}) for the relevant variables, assuming linear interactions between them. This approach uses the known form of entropy for Gaussian multivariates (\eq{gaussianEntropy}, in nats) \citep{cover91} and sums and differences of these entropies to compute other measures (e.g. transfer entropy as per \citep{kais02}). These estimators are fast ($\operatorname{O}\left(N d^2 \right)$, for dimensionality $d$ of the given joint variable) and parameter-free, but subject to the linear-model assumption.
	\item \textbf{Kernel estimation} of the relevant PDFs via a \textit{kernel function} are discussed in \app{boxKernelEstimators} (and see e.g. \citet{schr00}, \citet{kais02} and \citet{kantz97}). Such kernel functions measure similarity between pairs of samples using a specific resolution or \textit{kernel width} $r$; e.g. the box-kernel (implemented in JIDT) results in counting the proportion of the $N$ sample values which fall within $r$ of the given sample. They are then used as plug-in estimates for the entropy, and again sums and differences of these for the other measures.
Kernel estimation can measure non-linear relationships and is model-free (unlike Gaussian estimators), though is sensitive to the parameter choice for $r$ \citep{schr00,kais02} and is biased.
It is less time-efficient than the simple methods, although box-assisted methods can achieve $\operatorname{O}\left(N \right)$ time-complexity \citep{kantz97}.
See \app{boxKernelEstimators} for further comments, e.g. regarding selection of $r$.
	\item The \textbf{\citet*{kra04a}} (KSG) technique (see details in \app{ksgEstimators}) improved on (box-) kernel estimation for MI (and multi-information) via the use of Kozachenko-Leonenko estimators \citep{koza87} of log-probabilities via nearest-neighbour counting; bias correction; and a fixed number $K$ of nearest neighbours in the full $X$-$Y$ joint space. The latter effectively means using a dynamically altered (box-) kernel width $r$ to adjust to the density of samples in the vicinity of any given observation; this smooths out errors in the PDF estimation, especially when handling a small number of observations.
These authors proposed two slightly different algorithms for their estimator -- both are implemented in JIDT.
The KSG technique has been directly extended to conditional MI by \citet{fre07} and transfer entropy (originally by \citet{gom10a} and later for algorithm 2 by \citet{wib14c}).
KSG estimation builds on the non-linear and model-free capabilities of kernel estimation with bias correction, better data efficiency and accuracy, and being effectively parameter-free (being relatively stable to choice of $K$).
As such, it is widely-used as best of breed solution for MI, conditional MI and TE for continuous data; see e.g. \citet{wib14c} and \citet{vic14a}.
It can be computationally expensive with naive algorithms requiring $\operatorname{O}\left(K N^2 \right)$ time though fast nearest neighbour search techniques can reduce this to $\operatorname{O}\left(K N \log{N} \right)$.
For release v1.0 JIDT only implements a naive algorithm, though fast nearest neighbour search is implemented and available via the project SVN repository (see \secRef{installation}) and as such will be included in future releases.
	\item \textbf{Permutation entropy} approaches \citep{bandt02} estimate the relevant PDFs based on the relative ordinal structure of the joint vectors (see \app{permutationEstimators}). Permutation entropy has for example been adapted to estimate TE as the \textit{symbolic transfer entropy} \citep{stan08a}.
Permutation approaches are computationally fast, but are model-based however (assuming all relevant information is in the ordinal relationships).
This is not necessarily the case, and can lead to misleading results, as demonstrated by \citet{wib13a}.
\end{itemize}

\subsection{Related open-source information-theoretic toolkits}
\label{sec:otherToolkits}

We next consider other existing open-source information-theoretic toolkits for computing the aforementioned measures empirically from time-series data.
In particular we consider those which provide implementations of the transfer entropy.
For each toolkit, we describe its purpose, the type of data it handles, and which measures and estimators are implemented.

TRENTOOL (\href{http://www.trentool.de}{http://www.trentool.de}, GPL v3 license) by \citet{lind11a} is a MATLAB toolbox which is arguably the most mature open-source toolkit for computing TE.
It is not intended for general-purpose use, but designed from the ground up for transfer entropy analysis of (continuous) neural data, using the data format of the FieldTrip toolbox \citep{oost11} for EEG, MEG and LFP recordings.
In particular, it is designed for performing effective connectivity analysis between the input variables (see \citet{vic11a} and \citet{wib11a}), including statistical significance testing of TE results (as outlined in \app{statSigTesting}) and processing steps to deal with volume conduction and identify cascade or common-driver effects in the inferred network. Conditional/multivariate TE is not yet available, but planned.
TRENTOOL automates selection of parameters for embedding input time-series data and for source-target delays, and implements KSG estimation (see \app{ksgEstimators}), harnessing fast nearest neighbour search, parallel computation and GPU-based algorithms \citep{woll14a}.

The MuTE toolbox by \citet{mon14c,mon14a} (available via \href{http://figshare.com/articles/MuTE_toolbox_to_evaluate_Multivariate_Transfer_Entropy/1005245/1}{figshare}, CC-BY license \citep{mon14b}) provides MATLAB code for TE estimation.
In particular, MuTE is capable of computing conditional TE, includes a number of estimator types (discrete or binned, Gaussian, and KSG including fast nearest neighbour search), and adds non-uniform embedding (see \citet{faes11a}).
It also adds code to assist with embedding parameter selection, and incorporates statistical significance testing.

The Transfer entropy toolbox (TET, \href{http://code.google.com/p/transfer-entropy-toolbox/}{http://code.google.com/p/transfer-entropy-toolbox/}, BSD license) by \citet{ito11a} provides C-code callable from MATLAB for TE analysis of spiking data.
TET is limited to binary (discrete) data only.
Users can specify embedding dimension and source-target delay parameters.

MILCA (Mutual Information Least-dependent Component Analysis \href{http://www.ucl.ac.uk/ion/departments/sobell/Research/RLemon/MILCA/MILCA}{http://www.ucl.ac.uk/ion/departments/\-sobell/\-Research/RLemon/MILCA/MILCA}, GPL v3 license) provides C-code (callable from MATLAB) for mutual information calculations on continuous data \citep{ast13a,kra04a,stoeg04a}.
MILCA's purpose is to use the MI calculations as part of Independent Component Analysis (ICA), but they can be accessed in a general-purpose fashion.
MILCA implements KSG estimators with fast nearest neighbour search;
indeed, MILCA was co-written by the authors of this technique. It also handles multidimensional variables.

TIM (\href{http://www.cs.tut.fi/\%7etimhome/tim/tim.htm}{http://www.cs.tut.fi/\%7etimhome/tim/tim.htm}, GNU Lesser GPL license) by \citet{rut11a} provides C++ code (callable from MATLAB) for general-purpose calculation of a wide range of information-theoretic measures on continuous-valued time-series, including for multidimensional variables.
The measures implemented include entropy (Shannon, Renyi and Tsallis variants), Kullback-Leibler divergence, MI, conditional MI, TE and conditional TE.
TIM includes various estimators for these, including Kozachenko-Leonenko (see \app{ksgEstimators}), Nilsson-Kleijn \citep{nil07a} and Stowell-Plumbley \citep{stow09a} estimators for (differential) entropy, and KSG estimation for MI and conditional MI (using fast nearest neighbour search).

The MVGC (multivariate Granger causality toolbox, \href{http://www.sussex.ac.uk/sackler/mvgc/}{http://www.sussex.ac.uk/sackler/mvgc/}, GPL v3 license) by \citet{barn14a} provides a MATLAB implementation for general-purpose calculation of the Granger causality (i.e. TE with a linear-Gaussian model, see \app{infoDynamicsMeasures}) on continuous data.
MVGC also requires the MATLAB Statistics, Signal Processing and Control System Toolboxes.

There is a clear gap for a general-purpose information-theoretic toolkit, which can run in multiple code environments, implementing all of the measures in \app{classicMeasures} and \app{infoDynamicsMeasures}, with various types of estimators, and with implementation of local values, measures of statistical significance etc.
In the next section we introduce JIDT, and outline how it addresses this gap.
Users should make a judicious choice of which toolkit suits their requirements, taking into account data types, estimators and application domain.
For example, TRENTOOL is built from the ground up for effective network inference in neural imaging data, and is certainly the best tool for that application in comparison to a general-purpose toolkit.

\section{JIDT installation, contents and design}
\label{sec:toolkit}

JIDT (Java Information Dynamics Toolkit, \href{http://code.google.com/p/information-dynamics-toolkit/}{http://code.google.com/p/information-dynamics-toolkit/}, GPL v3 license) is unique as a general-purpose information-theoretic toolkit which provides all of the following features in one package:
\begin{itemize}
\item Implementation of a large array of measures, including all conditional/multivariate forms of the transfer entropy, complementary measures such as active information storage, and allows full specification of relevant embedding parameters;
\item Implementation a wide variety of estimator types and applicability to both discrete and continuous data;
\item Implementation of local measurement for all estimators;
\item Inclusion of statistical significance calculations for MI, TE, etc. and their conditional variants;
\item No dependencies on other installations (except Java).
\end{itemize}
Furthermore, JIDT is written in Java\footnote{The JIDT v1.0 distribution is compiled by Java Standard Edition 6; it is also verified as compatible with Edition 7.}, taking advantage of the following features:
\begin{itemize}
\item The code becomes platform agnostic, requiring only an installation of the Java Virtual Machine (JVM) to run;
\item The code is object-oriented, with common code shared and an intuitive hierarchical design using interfaces; this provides flexibility and allows different estimators of same measure can be swapped dynamically using polymorphism;
\item The code can be called directly from MATLAB, GNU Octave, Python, etc., but runs faster than native code in those languages (still slower but comparable to C/C++, see ``Computer Language Benchmarks Game''
\ifarXiv
	\citep{benchmark2014}); and
\else
	\citeyear{benchmark2014}); and
\fi
\item Automatic generation of Javadoc documents for each class.
\end{itemize}

In the following, we describe the (minimal) installation process in \secRef{installation}, and contents of the version 1.0 JIDT distribution in \secRef{contents}.
We then describe which estimators are implemented for each measure in \secRef{measuresImplemented}, and architecture of the source code in \secRef{architecture}.
We also outline how the code has been tested in \secRef{validation}, how to build it (if required) in \secRef{antScript} and point to other sources of documentation in \secRef{documentation}.

\begin{table*}[!t]
\caption{Relevant web/wiki pages on JIDT website.}
\label{table:wikiPages}
\centering
\ifarXiv\else\begin{small}\fi
\begin{tabular}{|c|l|}
 \hline
 Name & URL \\
 \hline
 Project home & \href{http://code.google.com/p/information-dynamics-toolkit/}{http://code.google.com/p/information-dynamics-toolkit/} \\
 Installation & \href{http://code.google.com/p/information-dynamics-toolkit/wiki/Installation}{http://code.google.com/p/information-dynamics-toolkit/wiki/Installation} \\
 Downloads & \href{http://code.google.com/p/information-dynamics-toolkit/wiki/Downloads}{http://code.google.com/p/information-dynamics-toolkit/wiki/Downloads} \\
 MATLAB/Octave use & \href{http://code.google.com/p/information-dynamics-toolkit/wiki/UseInOctaveMatlab}{http://code.google.com/p/information-dynamics-toolkit/wiki/UseInOctaveMatlab} \\
 Octave-Java array conversion & \href{http://code.google.com/p/information-dynamics-toolkit/wiki/OctaveJavaArrayConversion}{http://code.google.com/p/information-dynamics-toolkit/wiki/OctaveJavaArrayConversion} \\
 Python use & \href{http://code.google.com/p/information-dynamics-toolkit/wiki/UseInPython}{http://code.google.com/p/information-dynamics-toolkit/wiki/UseInPython} \\
  JUnit test cases & \href{http://code.google.com/p/information-dynamics-toolkit/wiki/JUnitTestCases}{http://code.google.com/p/information-dynamics-toolkit/wiki/JUnitTestCases} \\
  Documentation & \href{http://code.google.com/p/information-dynamics-toolkit/wiki/Documentation}{http://code.google.com/p/information-dynamics-toolkit/wiki/Documentation} \\
  Demos & \href{http://code.google.com/p/information-dynamics-toolkit/wiki/Demos}{http://code.google.com/p/information-dynamics-toolkit/wiki/Demos} \\
  Simple Java Examples & \href{http://code.google.com/p/information-dynamics-toolkit/wiki/SimpleJavaExamples}{http://code.google.com/p/information-dynamics-toolkit/wiki/SimpleJavaExamples} \\
  Octave/MATLAB Examples & \href{http://code.google.com/p/information-dynamics-toolkit/wiki/OctaveMatlabExamples}{http://code.google.com/p/information-dynamics-toolkit/wiki/OctaveMatlabExamples} \\
  Python Examples & \href{http://code.google.com/p/information-dynamics-toolkit/wiki/PythonExamples}{http://code.google.com/p/information-dynamics-toolkit/wiki/PythonExamples} \\
  Cellular Automata demos & \href{http://code.google.com/p/information-dynamics-toolkit/wiki/CellularAutomataDemos}{http://code.google.com/p/information-dynamics-toolkit/wiki/CellularAutomataDemos} \\
  Schreiber TE Demos & \href{http://code.google.com/p/information-dynamics-toolkit/wiki/SchreiberTeDemos}{http://code.google.com/p/information-dynamics-toolkit/wiki/SchreiberTeDemos} \\
  jidt-discuss group & \href{http://groups.google.com/group/jidt-discuss}{http://groups.google.com/group/jidt-discuss} \\
  SVN URL & \href{http://information-dynamics-toolkit.googlecode.com/svn/trunk/}{http://information-dynamics-toolkit.googlecode.com/svn/trunk/} \\
 \hline
\end{tabular}
\ifarXiv\else\end{small}\fi
\end{table*}

\subsection{Installation and dependencies}
\label{sec:installation}

There is \textbf{little to no installation} of JIDT required beyond downloading the software.
The software can be run on any platform which supports a standard edition Java Runtime Environment (i.e. Windows, Mac, Linux, Solaris).

Material pertaining to installation is described in full at the ``Installation'' wiki page for the project (\emph{see \tableRef{wikiPages} for all relevant project URLs}); summarised as follows:
\begin{enumerate}
	\item Download a code release package from the ``Downloads'' wiki page. Full distribution is recommended (described in \secRef{contents}) so as to obtain e.g. access to the examples described in \secRef{demos}, though a ``Jar only'' distribution provides just the JIDT library \texttt{infodynamics.jar} in Java archive file format.
	\item Unzip the full \texttt{.zip} distribution to the location of your choice, and/or move the \texttt{infodynamics.jar} file to a relevant location. Ensure that \texttt{infodynamics.jar} is on the Java classpath when your code attempts to access it (see \secRef{demos}).
	\item To update to a new version, simply copy the new distribution over the top of the previous one.
\end{enumerate}

As an alternative, advanced users can take an SVN checkout of the source tree from the SVN URL (see \tableRef{wikiPages}) and build the \texttt{infodynamics.jar} file using \texttt{ant} scripts (see \secRef{antScript}).

\textbf{In general, there are no dependencies} that a user would need to download in order to run the code. Some exceptions are as follows:
\begin{enumerate}
	\item\label{item:java} Java must be installed on your system in order to run JIDT; most systems will have Java already installed. To simply run JIDT, you will only need a Java Runtime Environment (JRE, also known as Java Virtual Machine or JVM), whereas to modify and/or build to software, or write your own Java code to access it, you will need the full Java Development Kit (JDK), standard edition (SE). Download it from \href{http://java.com/}{http://java.com/}. For using JIDT via MATLAB, a JVM is included in MATLAB already.
	\item If you wish to build the project using the \texttt{build.xml} script -- this requires \texttt{ant} (see \secRef{antScript}).
	\item If you wish to run the unit test cases (see \secRef{validation}) - this requires the \texttt{JUnit} framework: \href{http://www.junit.org/}{http://www.junit.org/} - for how to run \texttt{JUnit} with our ant script see ``JUnit test cases'' wiki page.
	\item Additional preparation may be required to use JIDT in GNU Octave or Python. Octave users must install the \texttt{octave-java} package from the \texttt{Octave-forge} project -- see description of these steps at ``MATLAB/Octave use'' wiki page. Python users must install a relevant Python-Java extension -- see description at ``Python use'' wiki page. Both cases will depend on a JVM on the system (as per point \ref{item:java} above), though the aforementioned extensions may install this for you. 
\end{enumerate}
Note that JIDT does \textit{adapt} code from a number of sources in accordance with their open-source license terms, including: Apache Commons Math v3.3 (\href{http://commons.apache.org/proper/commons-math/}{http://commons.apache.org/proper/commons-math/}), the JAMA project (\href{http://math.nist.gov/javanumerics/jama/}{http://math.nist.gov/javanumerics/jama/}), and the octave-java package from the Octave-Forge project (\href{http://octave.sourceforge.net/java/}{http://octave.\-sourceforge.net/java/}).
Relevant notices are supplied in the \texttt{notices} folder of the distribution.
Such code is included in JIDT however and does not need to be installed separately.

\subsection{Contents of distribution}
\label{sec:contents}

The contents of the current (version 1.0) JIDT (full) distribution are as follows:
\begin{itemize}
\item The top level folder contains the \texttt{infodynamics.jar} library file, a GNU GPL v3 license, a \texttt{readme.txt} file and an ant \texttt{build.xml} script for (re-)building the code (see \secRef{antScript});
\item The \texttt{java} folder contains source code for the library in the \texttt{source} subfolder (described in \secRef{measuresImplemented}), and unit tests in the \texttt{unittests} subfolder (see \secRef{validation}).
\item The \texttt{javadocs} folder contains automatically generated Javadocs from the source code, as discussed in \secRef{documentation}.
\item The \texttt{demos} folder contains several example applications of the software, described in \secRef{demos}, sorted into folders to indicate which environment they are intended to run in, i.e. \texttt{java}, \texttt{octave} (which is compatible with \texttt{MATLAB}) and \texttt{python}. There is also a \texttt{data} folder here containing sample data sets for these demos and unit tests.
\item The \texttt{notices} folder contains notices and licenses pertaining to derivations of other open source code used in this project.
\end{itemize}

\subsection{Source code and estimators implemented}
\label{sec:measuresImplemented}

The Java source code for the JIDT library contained in the \texttt{java/source} folder is organised into the following Java \textit{packages} (which map directly to subdirectories):
\begin{itemize}
\item \texttt{infodynamics.measures} contains all of the classes implementing the information-theoretic measures, split into:
	\begin{itemize}
		\item \texttt{infodynamics.measures.discrete} containing all of the measures for discrete data;
		\item \texttt{infodynamics.measures.continuous} which at the top level contains Java \textit{interfaces} for each of the measures as applied to continuous data, then a set of sub-packages (\texttt{gaussian}, \texttt{kernel}, \texttt{kozachenko}, \texttt{kraskov} and \texttt{symbolic}) which map to each estimator type in \secRef{estimationTechniques} and contain \textit{implementations} of such estimators for the interfaces defined for each measure (\secRef{architecture} describes the object-oriented design used here). \tableRef{implementedMeasures} identifies which estimators are measured for each estimator type;
		\item \texttt{infodynamics.measures.mixed} includes \textit{experimental} discrete-to-continuous MI calculators, though these are not discussed in detail here.
	\end{itemize}
	\item \texttt{infodynamics.utils} contains classes providing a large number of utility functions for the measures (e.g. matrix manipulation, file reading/writing including in Octave text format);
	\item \texttt{infodynamics.networkinference} contains implementations of higher-level algorithms which use the information-theoretic calculators to infer an effective network structure from time-series data (see \secRef{otherDemos}).
\end{itemize}

As outlined above, \tableRef{implementedMeasures} describes which estimators are implemented for each measure.
This effectively maps the definitions of the measures in \secRef{measures} to the estimators in \secRef{estimationTechniques} (note that the efficiency of these estimators is also discussed in \secRef{measures}).
All estimators provide the corresponding \textit{local} information-theoretic measures (as introduced in \app{localMeasures}).
Also, for the most part, the estimators include a generalisation to multivariate $\mathbf{X}$, $\mathbf{Y}$, etc, as identified in the table.

\begin{table*}[!t]
\caption{An outline of which estimation techniques are implemented for each relevant information-theoretic measure.
The \checkmark symbol indicates that the measure is implemented for the given estimator and is applicable to \textit{both} univariate and multivariate time-series (e.g. collective transfer entropy, where the source is multivariate), while the addition of superscript `$u$' (i.e. \checkmark$^u$) indicates the measure is implemented for univariate time-series only.
The section numbers and equation numbers refer to the definitions in the \appRef \ for the expected and local values (where provided) for each measure.
Also, while the KSG estimator is not applicable for entropy, the $\star$ symbol there indicates the implementation of the estimator by \citet{koza87} for entropy (which the KSG technique is based on for MI; see \app{ksgEstimators}).
Finally, $\dagger$ indicates that the (continuous) predictive information calculators are not available in the v1.0 release but are available via the project SVN and future releases.}
\label{table:implementedMeasures}
\centering
\ifarXiv\else\begin{footnotesize}\fi
\begin{tabular}{|c|c|c||c||c|c|c|c|}
 \hline
 \multicolumn{3}{|c||}{\textbf{Measure}} & Discrete & \multicolumn{4}{c|}{Continuous estimators} \\
 \cline{1-3} \cline{5-8} & & & estimator & Gaussian & Box-Kernel & Kraskov \etal (KSG) & Permutation \\
 Name & Notation & Defined at & \appRefAbbrev{discreteEstimator} & \appRefAbbrev{gaussianEstimators} & \appRefAbbrev{boxKernelEstimators} & \appRefAbbrev{ksgEstimators} & \appRefAbbrev{permutationEstimators} \\
 \hline
 \hline
 Entropy & $\entropy{X}$ & \eqs{entropy}{localEntropy} & \checkmark & \checkmark & \checkmark & $\star$ & \\
 \hline
\todo{Add conditional entropy here once it's implemented}
 Entropy rate & $\entropyRateCond{X}$ & \eqs{entropyRateConditionalKSTat}{localEntropyRateConditionalK} & \checkmark & \multicolumn{3}{c|}{\textit{Use two multivariate entropy calculators}} & \\
 \hline
 Mutual information (MI) & $\mutualInfo{X}{Y}$ & \eqs{mi}{localMi} & \checkmark & \checkmark & \checkmark & \checkmark \todo{Check -- website says this for MI and conditional MI, but I don't think it's uploaded yet. Response: they're uploaded for mixed calculator, and I'm not sure how else to represent that, so I'll leave it as is.} & \\
 \hline
 Conditional MI & $\condMutualInfo{X}{Y}{Z}$ & \eqs{condMi}{localCondMi} & \checkmark & \checkmark &  & \checkmark & \\
 \hline
 Multi-information & $\multiInfo{\mathbf{X}}$ & \eqs{multiInfo}{localMultiInfo} & \checkmark &  & \checkmark$^u$ & \checkmark$^u$ & \\
 \hline
 Transfer entropy (TE) & $\te{Y}{X}$ & \eqs{te}{localTEK} & \checkmark & \checkmark & \checkmark & \checkmark & \checkmark$^u$ \\
 \hline
 Conditional TE & $\condTe{Y}{X}{Z}$ & \eqs{teCondEstimate}{localCondTEK} & \checkmark & \checkmark$^u$ & & \checkmark$^u$ & \\
 \hline
 Active information storage & $\ais{X}$ & \eqs{activeStorageEstimate}{localActiveK} & \checkmark & \checkmark$^u$ & \checkmark$^u$ & \checkmark$^u$ & \\
 \hline
 Predictive information & $\excessEntropy{X}$ & \eqs{predictiveInfoFiniteK}{localPredictiveInfoFiniteK} & \checkmark & $\dagger^u$ & $\dagger^u$ & $\dagger^u$ & \\
 \hline
 Separable information & $\separable{X}$ & \eq{separableInfoEstimate} & \checkmark &  &  &  & \\
 \hline
\end{tabular}
\ifarXiv\else\end{footnotesize}\fi
\end{table*}

\subsection{JIDT architecture}
\label{sec:architecture}

\begin{figure*}[t]
	\begin{center}
		\label{fig:teKraskovClassDiagram}\includegraphics[trim= 20 540 30 20,clip=true,width=0.99\columnwidth]{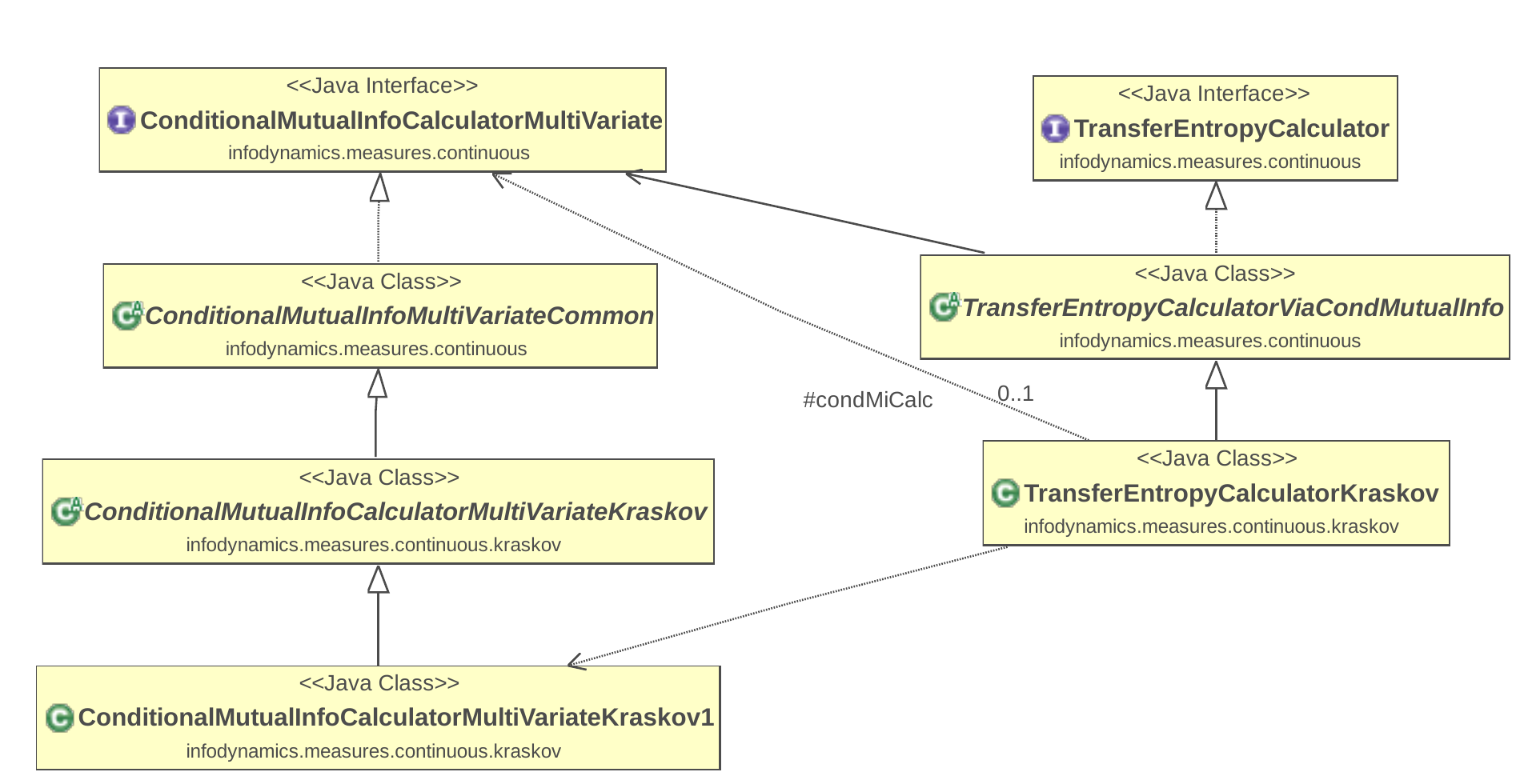}
	\end{center}
	\caption{ 
		  Partial UML class diagram of the implementations of the conditional mutual information (\eq{condMi}) and transfer entropy (\eq{te}) measures using KSG estimators. As explained in the main text, this diagram shows the typical \textit{object-oriented} structure of the implementations of various estimators for each measure. The relationships indicated on the class diagram are as follows: dotted lines with hollow triangular arrow heads indicate the realisation or \textit{implementation} of an interface by a class; solid lines with hollow triangular arrow heads indicate the generalisation or \textit{inheritance} of a child or subtype from a parent or superclass; lines with plain arrow heads indicate that one class \textit{uses} another (with the solid line indicating direct usage and dotted line indicating indirect usage via the superclass).
}
\end{figure*}

The measures for continuous data have been organised in a strongly \textit{object-oriented} fashion.\footnote{This is also the case for the measures for discrete data, though to a lesser degree and without multiple estimator types, so this is not focussed on here.}
\fig{teKraskovClassDiagram} provides a sample (partial) Unified Modeling Language (UML) class diagram of the implementations of the conditional mutual information (\eq{condMi}) and transfer entropy (\eq{te}) measures using KSG estimators (\app{ksgEstimators}).
This diagram shows the typical object-oriented hierarchical structure of the implementations of various estimators for each measure.
The class hierarchy is organised as follows.

\textbf{Interfaces} at the top layer define the available \textit{methods} for each measure.
At the top of this figure we see the  \texttt{ConditionalMutualInfoCalculatorMultiVariate} and \texttt{TransferEntropy\-Calculator} interfaces which define the methods each estimator \textit{class} for a given measure must implement. Such interfaces are defined for each information-theoretic measure in the \texttt{infodynamics.measures.\-continuous} package.

\textbf{Abstract classes}\footnote{Abstract classes provide implementations of some but not all methods required for a class, so they cannot be directly instantiated themselves but child classes which provide implementations for the missing methods and may be instantiated.} at the intermediate layer provide basic functionality for each measure.
Here, we have abstract classes \texttt{ConditionalMutualInfoMultiVariateCommon} and \texttt{TransferEntropy\-CalculatorViaCondMutualInfo} which \textit{implement} the above interfaces, providing common code bases for the given measures that various child classes can build on to specialise themselves to a particular estimator type. For instance, the \texttt{TransferEntropyCalculatorViaCond\-MutualInfo} class provides code which abstractly \textit{uses} a \texttt{ConditionalMutualInfoCalculatorMultiVariate} interface in order to make transfer entropy calculations, but does not concretely specify which type of conditional MI estimator to use, nor fully set its parameters.

\textbf{Child classes} at the lower layers add specialised functionality for each estimator type for each measure.
These child classes \textit{inherit} from the above parent classes, building on the common code base to add specialisation code for the given estimator type.
Here that is the KSG estimator type.
The child classes at the bottom of the hierarchy have no remaining abstract functionality, and can thus be used to make the appropriate information-theoretic calculation.
We see that \texttt{ConditionalMutualInfoCalculatorMultiVariateKraskov} begins to \textit{specialise} \texttt{ConditionalMutualInfoMultiVariateCommon} for KSG estimation, with further specialisation by its child class \texttt{ConditionalMutualInfoCalculatorMultiVariate\-Kraskov1} which implements the KSG algorithm 1 (\eq{kraskov1}). Not shown here is \texttt{ConditionalMutual\-InfoCalculatorMultiVariateKraskov2} which implements the KSG algorithm 2 (\eq{kraskov2}) and has similar class relationships.
We also see that \texttt{TransferEntropyCalculatorKraskov} specialises \texttt{TransferEntropyCalculatorViaCondMutualInfo} for KSG estimation, by using \texttt{ConditionalMutualInfoCalculator\-MultiVariateKraskov1} (or \texttt{ConditionalMutual\-InfoCalculatorMultiVariateKraskov2}, not shown) as the specific implementation of \texttt{ConditionalMutualInfoCalculatorMultiVariate}.
The implementations of these interfaces for other estimator types (e.g. \texttt{TransferEntropyCalculatorGaussian}) sit at the same level here inheriting from the common abstract classes above.

This type of object-oriented hierarchical structure delivers two important benefits: i. the \textit{decoupling} of common code away from specific estimator types and into common parent classes allows code re-use and simpler maintenance, and ii. the use of interfaces delivers \textit{subtype polymorphism} allowing \textit{dynamic dispatch}, meaning that one can write code to compute a given measure using the methods on its interface and only specify the estimator type at runtime (see a demonstration in \secRef{dynamicDispatch}).

\subsection{Validation}
\label{sec:validation}

The calculators in JIDT are validated using a set of \textit{unit tests} (distributed in the \texttt{java/unittests} folder).
Unit testing is a method of testing software by the use of a set of small test cases which call parts of the code and check the output against expected values, flagging errors if they arise.
The unit tests in JIDT are implemented via the \texttt{JUnit} framework version 3 (\href{http://www.junit.org/}{http://www.junit.org/}).
They can be run via the ant script (see \secRef{antScript}).

At a high level, the unit tests include validation of the results of information-theoretic calculations applied to the sample data in \texttt{demos/data} against measurements from various other existing toolkits, e.g.:
\begin{itemize}
	\item The KSG estimator (\app{ksgEstimators}) for MI is validated against values produced from the \textit{MILCA} toolkit \citep{ast13a,kra04a,stoeg04a};
	\item The KSG estimator for conditional MI and TE is validated against values produced from scripts within \textit{TRENTOOL} \citep{lind11a};
	\item The discrete and box-kernel estimators for TE are validated against the plots in the original paper on TE by \citet{schr00} (see \secRef{schreiberDemos});
	\item The Gaussian estimator for TE (\app{gaussianEstimators}) is verified against values produced from (a modified version of) the \texttt{computeGranger.m} script of the \textit{ChaLearn Connectomics Challenge Sample Code} \citep{chaLearnCode}.
\end{itemize}

Further code coverage by the unit tests is planned in future work.

\subsection{(Re-)building the code}
\label{sec:antScript}

Users may wish to build the code, perhaps if they are directly accessing the source files via \texttt{SVN} or modifying the files.
The source code may be compiled manually of course, or in your favourite IDE (Integrated Development Environment).
JIDT also provides an \texttt{ant} build script, \texttt{build.xml}, to guide and streamline this process.
Apache \texttt{ant} -- see \href{http://ant.apache.org/}{http://ant.apache.org/} -- is a command-line tool to build various interdependent targets in software projects, much like the older style \texttt{Makefile} for C/C++.

To build any of the following targets using \texttt{build.xml}, either integrate \texttt{build.xml} into your IDE and run the selected \texttt{<targetName>}, or run \texttt{ant <targetName>} from the command line in the top-level directory of the distribution, where \texttt{<targetName>} may be any of the following:
\begin{itemize}
	\item \texttt{build} or \texttt{jar} (this is the default if no \texttt{<targetName>} is supplied) -- creates a jar file for the JIDT library;
	\item \texttt{compile} -- compiles the JIDT library and unit tests;
	\item \texttt{junit} -- runs the unit tests;
	\item \texttt{javadocs} -- generates automated Javadocs from the formatted comments in the source code;
	\item \texttt{jardist} -- packages the JIDT jar file in a distributable form, as per the jar-only distributions of the project;
	\item \texttt{dist} -- runs unit tests, and packages the JIDT jar file, Javadocs, demos, etc. in a distributable form, as per the full distributions of the project;
	\item \texttt{clean} -- delete all compiled code etc. built by the above commands.
\end{itemize}

\subsection{Documentation and support}
\label{sec:documentation}

Documentation to guide users of JIDT is composed of:
\begin{enumerate}
	\item This manuscript!
	\item The Javadocs contained in the \texttt{javadocs} folder of the distribution (main page is \texttt{index.html}), and available online via the \texttt{Documentation} page of the project wiki (see \tableRef{wikiPages}). Javadocs are html formatted documentation for each package, class and interface in the library, which are automatically created from formatted comments in the source code. The Javadocs are very useful tools for users, since they provide specific details about each class and their methods, in more depth than we are able to do here; for example which properties may be set for each class. The Javadocs can be (re-)generated using ant as described in \secRef{antScript}.
	\item The demos; as described further in \secRef{demos}, on the Demos wiki page (see \tableRef{wikiPages}), and the individual wiki page for each demo;
	\item The project wiki pages (accessed from the project home page, see \tableRef{wikiPages}) provide additional information on various features, e.g. how to use JIDT in MATLAB or Octave and Python;
	\item The unit tests (as described in \secRef{validation}) provide additional examples on how to run the code.
\end{enumerate}

You can also join our email discussion group \texttt{jidt-discuss} on Google Groups (see URL in \tableRef{wikiPages}) or browse past messages, for announcements, asking questions, etc.

\section{JIDT code demonstrations}
\label{sec:demos}

In this section, we describe some simple demonstrations on how to use the JIDT library.
Several sets of demonstrations are included in the JIDT distribution, some of which are described here.
More detail is provided for each demo on its wiki page, accessible from the main Demos wiki page (see \tableRef{wikiPages}).
We begin with the main set of \textit{Simple Java Demos}, focussing in particular on a detailed walk-through of using a KSG estimator to compute transfer entropy since the calling pattern here is typical of all estimators for continuous data.
Subsequently, we provide more brief overviews of other examples available in the distribution, including how to run the code in MATLAB, GNU Octave and Python, implementing the transfer entropy examples from \citet{schr00}, and computing spatiotemporal profiles of information dynamics in Cellular Automata.

\subsection{Simple Java Demos}
\label{sec:simpleJavaDemos}

The primary set of demos is the ``Simple Java Demos'' set at \texttt{demos/java} in the distribution.
This set contains eight standalone Java programs to demonstrate simple use of various aspects of the toolkit.
This set is described further at the \texttt{SimpleJavaExamples} wiki page (see \tableRef{wikiPages}).

The Java source code for each program is located at \texttt{demos/java/infodynamics/demos} in the JIDT distribution, and shell scripts (with mirroring batch files for Windows\footnote{The batch files are not included in release v1.0, but are currently available via the SVN repository and will be distributed in future releases.}) to run each program are found at \texttt{demos/java/}.
The shell scripts demonstrate how to compile and run the programs from command line, e.g. \texttt{example1TeBinaryData.sh} contains the following commands:
\begin{lstlisting}[caption= {Shell script example1TeBinaryData.sh.},label={list:example1TeBinaryDataShell},name={example1TeBinaryDataShell},language={bash}]
# Make sure the latest source file is compiled.
javac -classpath "../../infodynamics.jar" "infodynamics/demos/Example1TeBinaryData.java"
# Run the example:
java -classpath ".:../../infodynamics.jar" infodynamics.demos.Example1TeBinaryData
\end{lstlisting}

The examples focus on various transfer entropy estimators (though similar calling paradigms can be applied to all estimators), including:
\begin{enumerate}
\item computing transfer entropy on \textit{binary (discrete) data};
\item computing transfer entropy for specific channels within \textit{multidimensional} binary data;
\item computing transfer entropy on \textit{continuous} data using \textit{kernel estimation};
\item computing transfer entropy on \textit{continuous} data using \textit{KSG estimation};
\item computing \textit{multivariate} transfer entropy on \textit{multidimensional} binary data;
\item computing \textit{mutual information} on continuous data, using \textit{dynamic dispatch} or \textit{late-binding} to a particular estimator;
\item computing transfer entropy from an \textit{ensemble} of time-series samples;
\item computing transfer entropy on \textit{continuous} data using \textit{binning} then \textit{discrete} calculation.
\end{enumerate}

In the following, we explore selected salient examples in this set.
We begin with \texttt{Example1TeBinary\-Data.java} and \texttt{Example4TeContinuousDataKraskov.java} as typical calling patterns to use estimators for discrete and continuous data respectively, then add extensions for how to compute local measures and statistical significance, use ensembles of samples, handle multivariate data and measures, and dynamic dispatch.

\subsubsection{Typical calling pattern for an information-theoretic measure on discrete data}
\label{sec:typicalDiscreteData}

\texttt{Example1Te\-BinaryData.java} (see \listRef{example1TeBinaryDataJava}) provides a typical calling pattern for calculators for \textbf{discrete data}, using the \texttt{infodynamics.\-measures.\-discrete.\-TransferEntropyCalculatorDiscrete} class.
\textit{While the specifics of some methods may be slightly different, the general calling paradigm is the same for all discrete calculators.}
\begin{lstlisting}[caption= {Estimation of TE from discrete data; source code adapted from \texttt{Example1TeBinaryData.java}.},label={list:example1TeBinaryDataJava},name={example1TeBinaryDataJava}]
int arrayLengths = 100;
RandomGenerator rg = new RandomGenerator();
// Generate some random binary data:
@\label{line:declarationIntArrays}@int[] sourceArray = rg.generateRandomInts(arrayLengths, 2);
@\label{line:declarationIntArrays2}@int[] destArray = new int[arrayLengths];
destArray[0] = 0;
System.arraycopy(sourceArray, 0, destArray, 1, arrayLengths - 1);
// Create a TE calculator and run it:
@\label{line:constructorDiscrete}@TransferEntropyCalculatorDiscrete teCalc = new TransferEntropyCalculatorDiscrete(2, 1);
@\label{line:initialiseDiscrete}@teCalc.initialise();
@\label{line:addObservationsDiscrete}@teCalc.addObservations(sourceArray, destArray);
@\label{line:computeDiscrete}@double result = teCalc.computeAverageLocalOfObservations();
\end{lstlisting}

The data type used for all discrete data are \texttt{int[]} time-series arrays (indexed by time).
Here we are computing TE for \textbf{univariate time series} data, so \texttt{sourceArray} and \texttt{destArray} at \lineRef{declarationIntArrays} and \lineRef{declarationIntArrays2} are single dimensional \texttt{int[]} arrays.
Multidimensional time series are discussed in \secRef{discreteJoint}.

The first step in using any of the estimators is to construct an instance of them, as per \lineRef{constructorDiscrete} above.
Parameters/properties for calculators for discrete data are \textit{only} supplied in the constructor at \lineRef{constructorDiscrete} (this is not the case for continuous estimators, see \secRef{typicalContinuousData}).
See the Javadocs for each calculator for descriptions of which parameters can be supplied in their constructor.
The arguments for the TE constructor here include: the number of discrete values ($M=2$), which means the data can take values $\{0,1\}$ (the allowable values are always enumerated $0,\ldots,M-1$); and the embedded history length $k=1$.
Note that for measures such as TE and AIS which require embeddings of time-series variables, the user must provide the embedding parameters here.

All calculators must be initialised before use or re-use on new data, as per the call to \texttt{initialise()} at \lineRef{initialiseDiscrete}.
This call clears any PDFs held inside the class, 
The \texttt{initialise()} method provides a mechanism by which the same object instance may be used to make separate calculations on multiple data sets, by calling it in between each application (i.e. looping from \lineRef{computeDiscrete} back to \lineRef{initialiseDiscrete} for a different data set -- see the full code for \texttt{Example1TeBinaryData.java} for an example).

The user then supplies the data to construct the PDFs with which the information-theoretic calculation is to be made.
Here, this occurs at \lineRef{addObservationsDiscrete} by calling the \texttt{addObservations()} method to supply the source and destination time series values.
This method can be called multiple times to add multiple sample time-series before the calculation is made (see further commentary for handling ensembles of samples in \secRef{multipleTrials}).

Finally, with all observations supplied to the estimator, the resulting transfer entropy may be computed via \texttt{computeAverageLocalOfObservations()} at \lineRef{computeDiscrete}.
The information-theoretic \textit{measurement is returned in bits} for all discrete calculators.
In this example, since the destination copies the previous value of the (randomised) source, then \texttt{result} should approach 1 bit.

\subsubsection{Typical calling pattern for an information-theoretic measure on continuous data}
\label{sec:typicalContinuousData}

Before outlining how to use the continuous estimators, we note that the discrete estimators above may be applied to continuous \texttt{double[]} data sets by first \textit{binning} them to convert them to \texttt{int[]} arrays, using either \texttt{MatrixUtils.discretise(double data[], int numBins)} for even bin sizes or \texttt{MatrixUtils.discretiseMaxEntropy( double data[], int numBins)} for maximum entropy binning (see \texttt{Example8TeContinuousDataByBinning}).
This is very efficient, however as per \secRef{estimationTechniques} it is more accurate to use an estimator which utilises the continuous nature of the data.

As such, we now review the use of a KSG estimator (\app{ksgEstimators}) to compute transfer entropy (\eq{te}), as a \textit{standard calling pattern} for all estimators applied to \textit{continuous data}.
The sample code in \listRef{ksgTransfer} is adapted from \texttt{Example4TeContinuousDataKraskov.java}.
(That this is a standard calling pattern can easily be seen by comparing to \texttt{Example3TeContinuousDataKernel.java}, which uses a box-kernel estimator but has very similar method calls, except for which parameters are passed in).
\newcounter{setObservationsStepNum}
\begin{lstlisting}[caption= {Use of KSG estimator to compute transfer entropy; adapted from \texttt{Example4TeContinuousDataKraskov.java}},label={list:ksgTransfer},name={ksgTransfer}]
@\label{line:declarationDoubleArrays}@double[] sourceArray, destArray;
// ...
// Import values into sourceArray and destArray
// ...
@\label{line:construct}@TransferEntropyCalculatorKraskov teCalc = new TransferEntropyCalculatorKraskov();
@\label{line:setProperty}@teCalc.setProperty("k", "4");
@\label{line:initialise}@teCalc.initialise(1);
@\setcounter{setObservationsStepNum}{\value{lstnumber}}
 \label{line:setObservations}@teCalc.setObservations(sourceArray, destArray);
@\label{line:compute}@double result = teCalc.computeAverageLocalOfObservations();
\end{lstlisting}

Notice that the calling pattern here is almost the same as that for discrete calculators, as seen in \listRef{example1TeBinaryDataJava}, with some minor differences outlined below.

Of course, for continuous data we now use \texttt{double[]} arrays (indexed by time) for the univariate time-series data here at \lineRef{declarationDoubleArrays}.
Multidimensional time series are discussed in \secRef{continuousJoint}.

As per discrete calculators, we begin by constructing an instance of the calculator, as per \lineRef{construct} above.
Here however, parameters for the operation of the estimator are not only supplied via the constructor (see below).
As such, all classes offer a constructor with no arguments, while only some implement constructors which accept certain parameters for the operation of the estimator.

Next, almost all relevant properties or parameters of the estimators can be supplied by passing key-value pairs of \texttt{String} objects to the \texttt{setProperty(String, String)} method at \lineRef{setProperty}.
The key values for properties which may be set for any given calculator are described in the Javadocs for the \texttt{setProperty} method for each calculator.
Properties for the estimator may be set by calling \texttt{setProperty} at any time; in most cases the new property value will take effect immediately, though it is only guaranteed to hold after the next initialisation (see below).
At \lineRef{setProperty}, we see that property \texttt{"k"} (shorthand for \texttt{ConditionalMutualInfoCalculatorMultiVariateKraskov.PROP\_K}) is set to the value \texttt{"4"}.
As described in the Javadocs for \texttt{TransferEntropyCalculatorKraskov.setProperty}, this sets the number of nearest neighbours $K$ to use in the KSG estimation in the full joint space.
Properties can also easily be extracted and set from a file, see \texttt{Example6LateBindingMutualInfo.java}.

As per the discrete calculators, all continuous calculators must be initialised before use or re-use on new data (see \lineRef{initialise}).
This clears any PDFs held inside the class, but additionally finalises any property settings here.
Also, the \texttt{initialise()} method for continuous estimators may accept some parameters for the calculator -- here it accepts a setting for the $k$ embedded history length parameter for the transfer entropy (see \eq{te}).
Indeed, there may be several \textit{overloaded} forms of \texttt{initialise()} for a given class, each accepting different sets of parameters.
For example, the \texttt{TransferEntropyCalculatorKraskov} used above offers an \texttt{initialise(k, tau\_k, l, tau\_l, u)} method taking arguments for both source and target embedding lengths $k$ and $l$, embedding delays $\tau_k$ and $\tau_l$ (see \app{infoDynamicsMeasures}), and source-target delay $u$ (see \eq{teDelay}). Note that currently such embedding parameters must be supplied by the user, although we intend to implement automated embedding parameter selection in the future.
Where a parameter is not supplied, the value given for it in a previous call to \texttt{initialise()} or \texttt{setProperty()} (or otherwise its default value) is used.

The supply of samples is also subtly different for continuous estimators.
Primarily, all estimators offer the \texttt{setObservations()} method (\lineRef{setObservations}) for supplying a single time-series of samples (which can only be done once).
See \secRef{multipleTrials} for how to use multiple time-series realisations to construct the PDFs via an \texttt{addObservations()} method.

Finally, the information-theoretic measurement (\lineRef{compute}) is returned in \textit{either bits or nats} as per the standard definition for this type of estimator in \secRef{estimationTechniques} (i.e. bits for discrete, kernel and permutation estimators; nats for Gaussian and KSG estimators).

At this point (before or after \lineRef{compute}) once all observations have been supplied, there are other quantities that the user may compute.
These are described in the next two subsections.

\subsubsection{Local information-theoretic measures}
\label{sec:jidtLocal}

\listRef{localCalculations} computes the \textit{local} transfer entropy (\eq{localTEK}) for the observations supplied earlier in \listRef{ksgTransfer}:
\begin{lstlisting}[caption= {Computing local measures after \listRef{ksgTransfer}; adapted from \texttt{Example4TeContinuousDataKraskov.java}.},label={list:localCalculations},name={ksgTransfer}]
@\label{line:computeLocal}@double[] localTE = teCalc.computeLocalOfPreviousObservations();
\end{lstlisting}
Each calculator (discrete or continuous) provides a \texttt{computeLocalOfPreviousObservations()} method to compute the relevant local quantities for the given measure (see \app{localMeasures}).
This method returns a \texttt{double[]} array of the local values (local TE here) at every time step $n$ for the supplied time-series observations.
For TE estimators, note that the first $k$ values (history embedding length) will have value zero, since local TE is not defined without the requisite history being available.\footnote{The return format is more complicated if the user has supplied observations via several \texttt{addObservations()} calls rather than \texttt{setObservations()}; see the Javadocs for \texttt{computeLocalOfPreviousObservations()} for details.}

\subsubsection{Null distribution and statistical significance}
\label{sec:jidtNullStatSig}

For the observations supplied earlier in \listRef{ksgTransfer}, \listRef{statCalculations} computes a distribution of surrogate TE values obtained via resampling under the null hypothesis that \texttt{sourceArray} and \texttt{destArray} have no temporal relationship (as described in \app{statSigTesting}).
\begin{lstlisting}[caption= {Computing null distribution after \listRef{ksgTransfer}; adapted from \texttt{Example3TeContinuousDataKernel.java}.},label={list:statCalculations},name={ksgTransfer}]
@\label{line:computeSignificance}@EmpiricalMeasurementDistribution dist = teCalc.computeSignificance(1000);
\end{lstlisting}
The method \texttt{computeSignificance()} is implemented for all MI and conditional MI based measures (including TE), for both discrete and continuous estimators.
It returns an \texttt{EmpiricalMeasurementDistribution} object, which contains a \texttt{double[]} array \texttt{distribut\-ion} of an empirical distribution of values obtained under the null hypothesis (the sample size for this distribution is specified by the argument to \texttt{computeSignificance()}).
The user can access the mean and standard deviation of the distribution, a $p$-value of whether these surrogate measurements were greater than the actual TE value for the supplied source, and a corresponding $t$-score (which assumes a Gaussian distribution of surrogate scores) via method calls on this object (see Javadocs for details).

Some calculators (discrete and Gaussian) overload the method \texttt{computeSignificance()} (without an input argument) to return an object encapsulating an \textit{analytically} determined $p$-value of surrogate distribution where this is possible for the given estimation type (see \app{statSigTesting}).
The availability of this method is indicated when the calculator implements the \texttt{AnalyticNullDistributionComputer} interface.

\subsubsection{Ensemble approach: using multiple trials or realisations to construct PDFs}
\label{sec:multipleTrials}

Now, the use of \texttt{setObservations()} for continuous estimators implies that the PDFs are computed from a single \textit{stationary} time-series realisation.
One may supply multiple time-series realisations (e.g. as multiple stationary trials from a brain-imaging experiment) via the following alternative calling pattern to \lineRef{setObservations} in \listRef{ksgTransfer}:
\let\oldthelstnumber=\thelstnumber
\renewcommand*\thelstnumber{\arabic{setObservationsStepNum}.\alph{lstnumber}}
\begin{lstlisting}[caption= {Supply of multiple time-series realisations as observations for the PDFs; an alternative to \lineRef{setObservations} in \listRef{ksgTransfer}. Code is adapted from \texttt{Example7EnsembleMethodTeContinuousDataKraskov.java}},label={list:addObservations}]
@\label{line:startAddObservations}@teCalc.startAddObservations();
@\label{line:addObservations1}@teCalc.addObservations(sourceArray1, destArray1);
@\label{line:addObservations2}@teCalc.addObservations(sourceArray2, destArray2);
@\label{line:addObservations3}@teCalc.addObservations(sourceArray3, destArray3);
// ...
@\label{line:finaliseAddObservations}@teCalc.finaliseAddObservations();
\end{lstlisting}
\let\thelstnumber=\oldthelstnumber
Computations on the PDFs constructed from this data can then follow as before.
Note that other variants of \texttt{addObservations()} exist, e.g. which pull out sub-sequences from the time series arguments; see the Javadocs for each calculator to see the options available.
Also, for the discrete estimators, \texttt{addObservations()} may be called multiple times directly without the use of a \texttt{startAddObservations()} or \texttt{finaliseAddObservations()} method.
This type of calling pattern may be used to realise an \textit{ensemble approach} to constructing the PDFs (see \citet{gom10a}; \citet{wib14c}; \citet{lind11a} and \citet{woll14a}), in particular by supplying only short corresponding (\textit{stationary}) parts of each trial to generate the PDFs for that section of an experiment.

\subsubsection{Joint-variable measures on multivariate discrete data}
\label{sec:discreteJoint}
For calculations involving \textbf{joint variables} from \textbf{multivariate discrete data} time-series (e.g. collective transfer entropy, see \app{infoDynamicsMeasures}), we use the \textit{same} discrete calculators (unlike the case for continuous-valued data in \secRef{continuousJoint}).
This is achieved with one simple pre-processing step, as demonstrated by \texttt{Example5TeBinaryMultivarTransfer.\-java}:
\begin{lstlisting}[caption= {Java source code adapted from \texttt{Example5TeBinaryMultivarTransfer.java}.},label={list:Example5TeBinaryMultivarTransfer},name={Example5TeBinaryMultivarTransfer}]
@\label{line:declarationInt2DArrays}@int[][] source, dest;
// ...
// Import binary values into the arrays,
//  with two columns each.
// ...
@\label{line:constructorDiscrete2}@TransferEntropyCalculatorDiscrete teCalc = new TransferEntropyCalculatorDiscrete(4, 1);
teCalc.initialise();
teCalc.addObservations(
@\label{line:preprocess1}@   MatrixUtils.computeCombinedValues(source, 2),
@\label{line:preprocess2}@   MatrixUtils.computeCombinedValues(dest, 2));
double result = teCalc.computeAverageLocalOfObservations();
\end{lstlisting}
We see that the multivariate discrete data is represented using two-dimensional \texttt{int[][]} arrays at \lineRef{declarationInt2DArrays}, where the first array index (row) is time and the second (column) is variable number.

The important pre-processing at \lineRef{preprocess1} and \lineRef{preprocess2} involves combining the joint vector of discrete values for each variable at each time step into a single discrete number; i.e. if our joint vector \texttt{source[t]} at time $t$ has $v$ variables, each with $M$ possible discrete values, then we can consider the joint vector as a $v$-digit base-$M$ number, and directly convert this into its decimal equivalent.
The \texttt{computeCombinedValues()} utility in \texttt{infodynamics.utils.MatrixUtils} performs this task for us at each time step, taking the \texttt{int[][]} array and the number of possible discrete values for each variable $M=2$ as arguments.
Note also that when the calculator was constructed at \lineRef{constructorDiscrete2}, we need to account for the total number of possible combined discrete values, being $M^v=4$ here.

\subsubsection{Joint-variable measures on multivariate continuous data}
\label{sec:continuousJoint}

For calculations involving \textbf{joint variables} from \textbf{multivariate continuous data} time-series, JIDT provides \textit{separate} calculators to be used.
\texttt{Example6LateBindingMutualInfo.java} demonstrates this for calculators implementing the \texttt{MutualInfoCalculatorMultiVariate} interface:\footnote{In fact, for MI, JIDT does not actually define a separate calculator for univariates -- the multivariate calculator \texttt{MutualInfoCalculatorMultiVariate} provides interfaces to supply univariate \texttt{double[]} data where each variable is univariate.}
\newcounter{instantiateStepNum}
\begin{lstlisting}[caption= {Java source code adapted from \texttt{Example6LateBindingMutualInfo.java}.},label={list:Example6LateBindingMutualInfo},name={Example6LateBindingMutualInfo}]
@\label{line:declarationDoubleArrays2}@double[][] variable1, variable2;
@\label{line:declarationMICalc}@MutualInfoCalculatorMultiVariate miCalc;
// ...
// Import continuous values into the arrays
@\setcounter{instantiateStepNum}{\value{lstnumber}}
\label{line:instantiateMiCalc}@//  and instantiate miCalc
// ...
@\label{line:jointMiInitialisation}@miCalc.initialise(2, 2);
miCalc.setObservations(variable1, variable2);
@\label{line:jointMiCalculation}@double miValue = miCalc.computeAverageLocalOfObservations();
\end{lstlisting}
First, we see that the multivariate continuous data is represented using two-dimensional \texttt{double[][]} arrays at \lineRef{declarationDoubleArrays2}, where  (as per \secRef{discreteJoint}) the first array index (row) is time and the second (column) is variable number.
The instantiating of a class implementing the \texttt{MutualInfoCalculatorMultiVariate} interface to make the calculations is not shown here (but is discussed separately in \secRef{dynamicDispatch}).

Now, a crucial step in using the multivariate calculators is specifying in the arguments to \texttt{initialise()} the number of dimensions (i.e. the number of variables or columns) for each variable involved in the calculation.
At \lineRef{jointMiInitialisation} we see that each variable in the MI calculation has two dimensions (i.e. there will be two columns in each of \texttt{variable1} and \texttt{variable2}).

Other interactions with these multivariate calculators follow the same form as for the univariate calculators.

\subsubsection{Coding to interfaces; or dynamic dispatch}
\label{sec:dynamicDispatch}

\listRef{Example6LateBindingMutualInfo} (\texttt{Example6LateBindingMutualInfo.\-java}) also demonstrates the manner in which a user can write code to use the \textit{interfaces} defined in \texttt{infodynamics.measures.continuous} -- rather than any particular \textit{class} implementing that measure -- and dynamically alter the instantiated class implementing this interface at runtime.
This is known as \textit{dynamic dispatch}, enabled by the \textit{polymorphism} provided by the interface (described at \secRef{architecture}).
This is a useful feature in object-oriented programming where, here, a user wishes to write code which requires a particular measure, and dynamically switch-in different estimators for that measure at runtime.
For example, in \listRef{Example6LateBindingMutualInfo} we may normally use a KSG estimator, but switch-in a linear-Gaussian estimator if we happen to know our data is Gaussian.

To use dynamic dispatch with JIDT:
\begin{enumerate}
	\item Write code to use an interface for a calculator (e.g. \texttt{MutualInfoCalculatorMultiVariate} in \listRef{Example6LateBindingMutualInfo}), rather than to directly use a particular implementing class (e.g. \texttt{MutualInfoCalculator\-MultiVariateKraskov});
	\item Instantiate the calculator object by \textit{dynamically} specifying the implementing class (compare to the static instantiation at \lineRef{construct} of \listRef{ksgTransfer}), e.g. using a variable name for the class as shown in \listRef{dynamicInstantiation}:
\end{enumerate}
\let\oldthelstnumber=\thelstnumber
\renewcommand*\thelstnumber{\arabic{instantiateStepNum}.\alph{lstnumber}}
\begin{lstlisting}[caption= {Dynamic instantiation of a mutual information calculator, belonging at \lineRef{instantiateMiCalc} in \listRef{Example6LateBindingMutualInfo}. Adapted from \texttt{Example6LateBindingMutualInfo.java}.},label={list:dynamicInstantiation}]
String implementingClass;
// Load the name of the class to be used into the
//   variable implementingClass
@\label{line:dynInstantiationLine}@miCalc = (MutualInfoCalculatorMultiVariate) Class.forName(implementingClass).newInstance();
\end{lstlisting}
\let\thelstnumber=\oldthelstnumber
Of course, to be truly dynamic, the value of \texttt{implementingClass} should not be hard-coded but must be somehow set by the user.
For example, in the full \texttt{Example6LateBindingMutualInfo.java} it is set from a properties file.

\subsection{MATLAB / Octave Demos}
\label{sec:matlabDemos}

The ``Octave/MATLAB code examples'' set at \texttt{demos/octave} in the distribution provide a basic set of demonstration scripts for using the toolkit in GNU Octave or MATLAB.
The set is described in some detail at the \texttt{OctaveMatlabExamples} wiki page (\tableRef{wikiPages}).
See \secRef{installation} regarding installation requirements for running the toolkit in Octave, with more details at the \texttt{UseInOctaveMatlab} wiki page (see \tableRef{wikiPages}).

The scripts in this set mirror the Java code in the ``Simple Java Demos'' set (\secRef{simpleJavaDemos}), to demonstrate that \textit{anything which JIDT can do in a Java environment can also be done in MATLAB/Octave}.
The user is referred to the distribution or the \texttt{OctaveMatlabExamples} wiki page for more details on the examples.
An illustrative example is provided in \listRef{example1TeBinaryDataMatlab}, which converts \listRef{example1TeBinaryDataJava} into MATLAB/Octave:
\begin{lstlisting}[caption= {Estimation of TE from discrete data in MATLAB/Octave; adapted from \texttt{example1TeBinaryData.m}},label={list:example1TeBinaryDataMatlab},name={example1TeBinaryDataMatlab}]
@\label{line:javaaddpath}@javaaddpath('../../infodynamics.jar');
sourceArray=(rand(100,1)>0.5)*1; 
destArray = [0; sourceArray(1:99)];
@\label{line:matlabConstructor}@teCalc=javaObject('infodynamics.measures.discrete.TransferEntropyCalculatorDiscrete', 2, 1);
teCalc.initialise();
teCalc.addObservations(
@\label{line:octaveArrayConversion1}@    octaveToJavaIntArray(sourceArray),
@\label{line:octaveArrayConversion2}@    octaveToJavaIntArray(destArray));
result = teCalc.computeAverageLocalOfObservations()
\end{lstlisting}
This example illustrates several important steps for using JIDT from a MATLAB/Octave environment:
\begin{enumerate}
	\item Specify the classpath (i.e. the location of the \texttt{infodynamics.jar} library) before using JIDT with the function \texttt{javaaddpath(samplePath)} (at \lineRef{javaaddpath});
	\item Construct classes using the \texttt{javaObject()} function (see \lineRef{matlabConstructor});
	\item Use of objects is otherwise almost the same as in Java itself, however:
	\begin{enumerate}
		\item In Octave, conversion between native \textit{array} data types and Java arrays is not straightforward; we recommend using the supplied functions for such conversion in \texttt{demos/octave}, e.g. \texttt{octaveToJavaIntArray.m}. These are described on the \texttt{OctaveJavaArrayConversion} wiki page (\tableRef{wikiPages}), and see example use in \lineRef{octaveArrayConversion1} here, and in \texttt{example2TeMultidimBinary\-Data.m} and \texttt{example5TeBinaryMultivarTransfer.m}.
		\item In Java arrays are indexed from 0, whereas in Octave or MATLAB these are indexed from 1. So when you call a method on a Java object such as \texttt{MatrixUtils.select(double data, int fromIndex, int length)} -- even from within MATLAB/Octave -- you must be aware that \texttt{fromIndex} will be indexed from 0 inside the toolkit, not 1!
	\end{enumerate}
\end{enumerate}

\subsection{Python Demos}
\label{sec:pythonDemos}

Similarly, the ``Python code examples'' set at \texttt{demos/python} in the distribution provide a basic set of demonstration scripts for using the toolkit in Python.
The set is described in some detail at the \texttt{PythonExamples} wiki page (\tableRef{wikiPages}).
See \secRef{installation} regarding installation requirements for running the toolkit in Python, with more details at the \texttt{UseInPython} wiki page (\tableRef{wikiPages}).

Again, the scripts in this set mirror the Java code in the ``Simple Java Demos'' set (\secRef{simpleJavaDemos}), to demonstrate that \textit{anything which JIDT can do in a Java environment can also be done in Python}.

Note that this set uses the \texttt{JPype} library (\href{http://jpype.sourceforge.net/}{http://jpype.sourceforge.net/}) to create the Python-Java interface, and the examples would need to be altered if you wish to use a different interface.
The user is referred to the distribution or the \texttt{PythonExamples} wiki page for more details on the examples.

An illustrative example is provided in \listRef{example1TeBinaryDataPython}, which converts \listRef{example1TeBinaryDataJava} into Python:
\begin{lstlisting}[caption= {Estimation of TE from discrete data in Python; adapted from \texttt{example1TeBinaryData.py}},label={list:example1TeBinaryDataPython},name={example1TeBinaryDataPython}]
@\label{line:jpypeImport}@from jpype import *
import random
@\label{line:startJVM}@startJVM(getDefaultJVMPath(), "-ea", "-Djava.class.path=../../infodynamics.jar")
sourceArray = [random.randint(0,1) for r in xrange(100)]
destArray = [0] + sourceArray[0:99];
@\label{line:pythonConstruct1}@teCalcClass = JPackage("infodynamics.measures.discrete").TransferEntropyCalculatorDiscrete
@\label{line:pythonConstruct2}@teCalc = teCalcClass(2,1)
teCalc.initialise()
teCalc.addObservations(sourceArray, destArray)
result = teCalc.computeAverageLocalOfObservations()
@\label{line:pythonShutdownJVM}@shutdownJVM() 
\end{lstlisting}
This example illustrates several important steps for using JIDT from Python via JPype:
\begin{enumerate}
	\item Import the relevant packages from JPype (\lineRef{jpypeImport});
	\item Start the JVM and specify the classpath (i.e. the location of the \texttt{infodynamics.jar} library) before using JIDT with the function \texttt{startJVM()} (at \lineRef{startJVM});
	\item Construct classes using a reference to their package (see \lineRef{pythonConstruct1} and \lineRef{pythonConstruct2});
	\item Use of objects is otherwise almost the same as in Java itself, however conversion between native \textit{array} data types and Java arrays can be tricky -- see comments on the \texttt{UseInPython} wiki page (see \tableRef{wikiPages}).
	\item Shutdown the JVM when finished (\lineRef{pythonShutdownJVM}).
\end{enumerate}

\subsection{Schreiber's Transfer Entropy Demos}
\label{sec:schreiberDemos}

The ``Schreiber Transfer Entropy Demos'' set at \texttt{demos/octave/SchreiberTransferEntropy\-Examples} in the distribution recreates the original examples introducing transfer entropy by \citet{schr00}.
The set is described in some detail at the \texttt{SchreiberTeDemos} wiki page (see \tableRef{wikiPages}).
The demo can be run in MATLAB or Octave.

The set includes computing TE with a discrete estimator for data from a Tent Map simulation, with a box-kernel estimator for data from a Ulam Map simulation, and again with a box-kernel estimator for heart and breath rate data from a sleep apnea patient\footnote{This data set was made available via the Santa Fe Institute time series contest held in 1991 \citep{rig93} and redistributed with JIDT with kind permission from Andreas Weigend.}  (see \citet{schr00} for further details on all of these examples and map types).
Importantly, the demo shows correct values for important parameter settings (e.g. use of bias correction) which were not made clear in the original paper.

\begin{figure}[t!]
  \begin{center}
		\includegraphics[width=0.5\columnwidth]{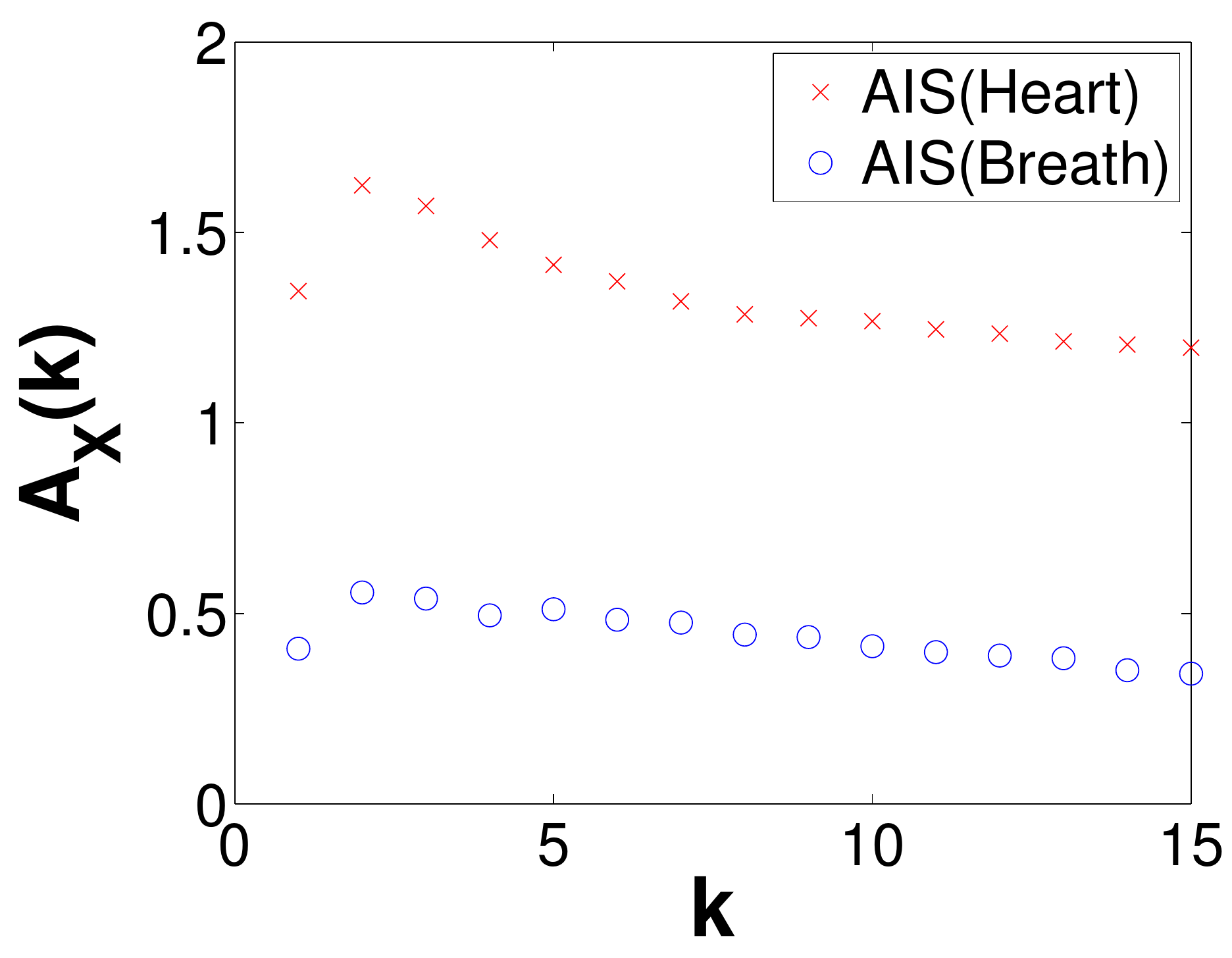}
	\end{center}
	\caption{\label{fig:heartBreathAIS} Active information storage (AIS) computed by the KSG estimator ($K=4$ nearest neighbours) as a function of embedded history length $k$ for the heart and breath rate time-series data.
	}
\end{figure}

We also revisit the heart-breath rate analysis using a KSG estimator, demonstrating how to select embedding dimensions $k$ and $l$ for this data set.
As an example, we show in \fig{heartBreathAIS} a calculation of AIS (\eq{activeStorageEstimate}) for the heart and breath rate data, using a KSG estimator with $K=4$ nearest neighbours, as a function of embedding length $k$.
This plot is produced by calling the MATLAB function: \texttt{activeInfoStorageHeartBreathRatesKraskov(1:15, 4)}.
Ordinarily, as an MI the AIS will be non-decreasing with $k$, while an observed increase may be simply because bias in the underlying estimator increases with $k$ (as the statistical power of the estimator is exhausted).
This is not the case however when we use an underlying KSG estimator, since the bias is automatically subtracted away from the result.
As such, we can use the peak of this plot to suggest that an embedded history of $k=2$ for both heart and breath time-series is appropriate to capture all relevant information from the past without adding more spurious than relevant information as $k$ increases.
(The result is stable with the number of nearest neighbours $K$.)
We then continue on to use those embedding lengths for further investigation with the TE in the demonstration code.

\subsection{Cellular Automata Demos}
\label{sec:caDemos}

The ``Cellular Automata Demos'' set at \texttt{demos/octave/CellularAutomata} in the distribution provide a standalone demonstration of the utility of \textit{local} information dynamics profiles.
The scripts allow the user to reproduce the key results from \citet{liz08a,liz10e,liz12a,liz14a}; \citet{liz13a}; \citet{liz13c} etc., i.e. plotting local information dynamics measures at every point in space-time in the cellular automata (CA).
These results confirmed the long-held conjectures that gliders are the dominant information transfer entities in CAs, while blinkers and background domains are the dominant information storage components, and glider/particle collisions are the dominant information modification events.

\begin{figure}[t!]
  \begin{center}
		\subfigure[Raw CA]{\label{fig:54-raw}\makebox[\caWidth]{\includegraphics[trim= 0 0 95 0,clip=true,height=\caHeight]{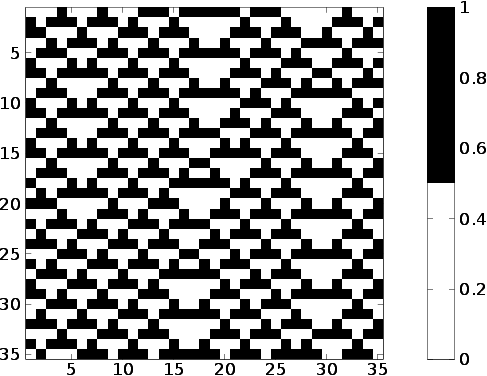}\ \ \ \ \ \ }} \ \ 
		\subfigure[$\localAis{X}{n,k=16}$]{\label{fig:54-active-colour-16}\makebox[\caWidth]{\includegraphics[height=\caHeight]{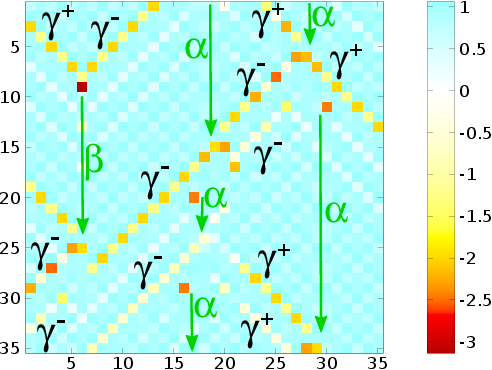}}}
		\subfigure[$\localTe{Y}{X}{n,k=16}$ right]{\label{fig:54-te-1-colour-16}\makebox[\caWidth]{\includegraphics[height=\caHeight]{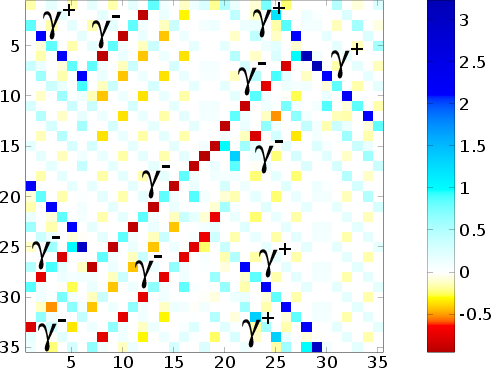}}} \ \ 
		\subfigure[$\localTe{Y}{X}{n,k=16}$ left]{\label{fig:54-te--1-colour-16}\makebox[\caWidth]{\includegraphics[height=\caHeight]{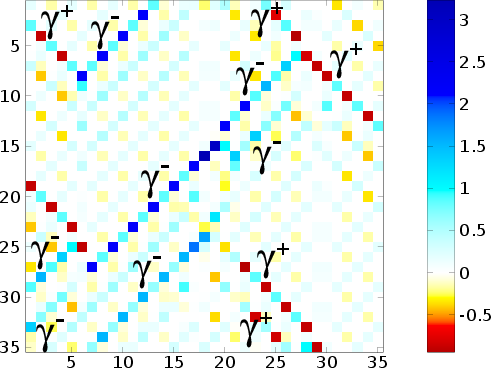}}}
	\end{center}
	\caption[Local information dynamics in elementary CA rule 54]{\label{fig:54} Local information dynamics in ECA \textbf{rule 54} for the raw values in \subref{fig:54-raw} (black for ``1'', white for ``0''). 35 time steps are displayed for 35 cells, and time increases down the page for all CA plots. All units are in bits, as per scales on the right-hand sides.
		\subref{fig:54-active-colour-16} Local active information storage; 
		Local apparent transfer entropy: \subref{fig:54-te-1-colour-16} one cell to the right, and 
		\subref{fig:54-te--1-colour-16} one cell to the left per time step. 
		NB: Reprinted with kind permission of Springer Science+Business Media from \citet{liz14b}, in \emph{Directed Information Measures in Neuroscience}, Understanding Complex Systems, edited by M. Wibral, R. Vicente, and J. T. Lizier (Springer, Berlin/Heidelberg, 2014) pp. 161--193.
	}
\end{figure}

The set is described in some detail at the \texttt{CellularAutomataDemos} wiki page (see \tableRef{wikiPages}).
The demo can be run in MATLAB or Octave.
The main file for the demo is \texttt{plotLocalInfoMeasureForCA.m}, which can be used to specify a CA type to run and which measure to plot an information profile for.
Several higher-level scripts are available to demonstrate how to call this, including \texttt{DirectedMeasuresChapterDemo2013.m} which was used to generate the figures by \citet{liz14b} (reproduced in \fig{54}).

\subsection{Other Demos}
\label{sec:otherDemos}

The toolkit contains a number of other demonstrations, which we briefly mention here:
\begin{itemize}
	\item The ``Interregional Transfer demo'' set at \texttt{demos/java/interregionalTransfer/} is a higher-level example of computing information transfer between two regions of variables (e.g. brain regions in fMRI data), using multivariate extensions to the transfer entropy, to infer effective connections between the regions. This demonstration implements the method originally described by \citet{liz11a}. Further documentation is provided via the \texttt{Demos} wiki page (see \tableRef{wikiPages}).
	\item The ``Detecting interaction lags'' demo set at \texttt{demos/octave/DetectingInteractionLags} shows how to use the transfer entropy calculators to investigate a source-destination lag that is different to 1 (the default). In particular, this demo was used to make the comparisons of using transfer entropy (TE) and momentary information transfer (MIT) \citep{pomp11} to investigate source-destination lags by \citet{wib13a} (see Test cases Ia and Ib therein). In particular, the results show that TE is most suitable for investigating source-destination lags as MIT can be deceived by source memory, and also that symbolic TE (\app{permutationEstimators}) can miss important components of information in an interaction. Further documentation is provided via the \texttt{Demos} wiki page (see \tableRef{wikiPages}).
	\item The ``Null distribution'' demo set at \texttt{demos/octave/NullDistributions} explores the match between analytic and resampled distributions of MI, conditional MI and TE under null hypotheses of no relationship between the data sets (see \app{statSigTesting}). Further documentation is provided via the \texttt{Demos} wiki page (see \tableRef{wikiPages}).
\end{itemize}

Finally, we note that demonstrations on using JIDT within several additional languages (Julia, Clojure and R) are currently available within the SVN repository only, and will be distributed in future releases.

\todo{ THINK ABOUT INCLUDING SOME / ALL OF THESE:

Practical considerations:
Importance of stationarity when being evaluated from one time-series.
Should be a representative time-series.
Which estimator to use?
}

\section{Conclusion}
\label{sec:conclusion}

We have described the Java Information Dynamics Toolkit (JIDT), an open-source toolkit available on Google code, which implements information-theoretic measures of dynamics via several different estimators.
We have described the architecture behind the toolkit and how to use it, providing several detailed code demonstrations.

In comparison to related toolkits, JIDT provides implementations for a wider array of information-theoretic measures, with a wider variety of estimators implemented, adds implementations of local measures and statistical significance, and is standalone software.
Furthermore, being implemented in Java, JIDT is platform agnostic and requires little to no installation, is fast, exhibits an intuitive object-oriented design, and can be used in MATLAB, Octave, Python and other environments.

JIDT has been used to produce results in publications by both this author and others \citep{liz13c,liz14a,liz14b,wib13a,wib14a,wib14b,gomez14a,wang14a,das13a,wang12a}.

It may be complemented by the Java Partial Information Decomposition (JPID) toolkit \citep{liz12e,liz13b}, which implements early attempts \citep{will10a} to separately measure redundant and synergistic components of the conditional mutual information (see \app{classicMeasures}).

We are planning the extension or addition of several important components in the future.
Of highest priority: we are exploring the use of multi-threading and GPU computing, and automated parameter selection for time-series embedding.
\todo{Should we mention collaborating with developers of TRENTOOL ??}
We will add additional implementations to complete \tableRef{implementedMeasures}, and aim for larger code coverage by our unit tests.
Most importantly however, we seek collaboration on the project from other developers in order to expand the capabilities of JIDT, and we will welcome volunteers who wish to contribute to the project.

\ifarXiv
\begin{acknowledgments}
\else
\section*{Disclosure/Conflict-of-Interest Statement}
The authors declare that the research was conducted in the absence of any commercial or financial relationships that could be construed as a potential conflict of interest.
\section*{Acknowledgement}
\fi
\label{sec:acks}
I am grateful to many collaborators and contacts who have tested JIDT in their applications, provided test data, found bugs, provided feedback, and reviewed this manuscript etc.
These include primarily: X. Rosalind Wang, Michael Wibral, Oliver Cliff, 
Siddharth Pritam, Rommel Ceguerra, Ipek \"Ozdemir
and Oliver Obst; 
as well as: Sakyasingha Dasgupta, Heni Ben Amor, Mizuki Oka, Christoph Hartmann, Michael Harr\'e and Patricia Wollstadt.
\ifarXiv
\end{acknowledgments}
\else
\bibliographystyle{frontiersinSCNS&ENG} 
\fi

\bibliography{references}

\ifarXiv
\appendix
\input{SupplementaryMaterialContents}
\else
\fi

\end{document}

%% file: SupplementaryMaterialContents.tex
\section{Information-theoretic measures}
\label{app:measures}

In this section, we give an overview of the information-theoretic measures which are implemented in JIDT.
We begin by describing basic information-theoretic measures such as entropy and mutual information in \appRefSec{classicMeasures}, then go on to describe in \appRefSec{infoDynamicsMeasures} the more contemporary measures which are being used to quantify the information dynamics of distributed computation.
The latter are the real focus of the toolkit.
We also describe in \appRefSec{localMeasures} how one can measure local or pointwise information-theoretic measures (to assign information values to specific observations or outcomes of variables and their interactions), the extension of the measures to continuous variables in \appRefSec{differentialEntropy}, and in \appRefSec{statSigTesting} how one can evaluate the statistical significance of the interaction between variables.
All features discussed are available in JIDT unless otherwise noted.

\subsection{Basic information-theoretic measures}
\label{app:classicMeasures}

We first outline basic information-theoretic measures \citep{cover91,mac03} implemented in JIDT.

The fundamental quantity of information theory is the \textbf{Shannon entropy}, which represents the expected or average uncertainty associated with any measurement $x$ of a random variable $X$:\footnote{Notation for all quantities is summarised in \tableRef{measuresSummary}.}
\begin{align}
	\entropy{X} = -\sum_{x \in \alpha_x} p(x) \log_2{p(x)}
	\label{eq:entropy}.
\end{align}
with a probabilities distribution function $p$ defined over the alphabet $\alpha_x$ of possible outcomes for $x$ (where $\alpha_x = \{0,\ldots,M_X-1\}$ without loss of generality for some $M_X$ discrete symbols).
Note that unless otherwise stated, logarithms are taken by convention in base 2, giving units in bits.

The Shannon entropy was originally derived following an axiomatic approach, being derived as the unique formulation (up to the base of the logarithm) satisfying a certain set of properties or axioms (see \citet{shan48} for further details).
The uncertainty $\entropy{X}$ associated with a measurement of $X$ is equal to the expected information required to predict it (see self-information below).
$\entropy{X}$ for a measurement $x$ of $X$ can also be interpreted as the minimal expected or average number of bits required to encode or describe its value without losing information \citep{mac03,cover91}.

The \textbf{joint entropy} of two random variables $X$ and $Y$ is a generalization to quantify the uncertainty of their joint distribution:
\begin{align}
	\jointEntropy{X}{Y} = -\sum_{x \in \alpha_x}{\sum_{y \in \alpha_y}{ p(x,y) \log_2{p(x,y)}}}
	\label{eq:jointEntropy}.
\end{align}
We can of course write the above equation for multivariate $\mathbf{Z} = \{X,Y\}$, and then generalise to $\entropy{\mathbf{X}}$ for $\mathbf{X} = \left\{ X_1,X_2,\ldots,X_G \right\}$. Such expressions for entropies of multivariates allows us to expand \textit{all} of the following quantities for multivariate $\mathbf{X}$, $\mathbf{Y}$ etc.

The \textbf{conditional entropy} of $X$ given $Y$ is the expected uncertainty that remains about $x$ when $y$ is known:
\begin{align}
	\condEntropy{X}{Y} = -\sum_{x \in \alpha_x}{\sum_{y \in \alpha_y}{ p(x,y) \log_2{p(x \mid y)}}}
	\label{eq:conditionalEntropy}.
\end{align}
The conditional entropy for a measurement $x$ of $X$ can be interpreted as the minimal expected number of bits required to encode or describe its value without losing information, given that the receiver of the encoding already knows the value $y$ of $Y$.
The previous quantities are related by the following \emph{chain rule}:
\begin{align}
	\jointEntropy{X}{Y} = \entropy{X} + \condEntropy{Y}{X}
	\label{eq:entropyChainRule}.
\end{align}

The \textbf{mutual information} (MI) between $X$ and $Y$ measures the expected reduction in uncertainty about $x$ that results from learning the value of $y$, or vice versa: 
\begin{align}
	\mutualInfo{X}{Y} & = \sum_{x \in \alpha_x}{\sum_{y \in \alpha_y}{ p(x,y) \log_2{\frac{p(x \mid y)}{p(x)}} }} \\
		& = \entropy{X} - \condEntropy{X}{Y}
	\label{eq:mi}.
\end{align}
The MI is symmetric in the variables $X$ and $Y$.
The mutual information for measurements $x$ and $y$ of $X$ and $Y$ can be interpreted as the expected number of bits \emph{saved} in encoding or describing $x$ given that the receiver of the encoding already knows the value of $y$, in comparison to the encoding of $x$ without the knowledge of $y$. These descriptions of $x$ with and without the value of $y$ are both minimal without losing information.
Note that one can compute the \emph{self-information} $\mutualInfo{X}{X} = \entropy{X}$.
Finally, one may define a generalization of the MI to a set of more than two variables $\mathbf{X} = \left\{ X_1,X_2,\ldots,X_G \right\}$, known as the \textbf{multi-information} or \textbf{integration} \citep{ton94}:
\begin{align}
	\multiInfo{\mathbf{X}} &= \multiInfoPlus{X_1}{X_2}{X_G} \nonumber \\
	&= \left( \sum_{g=1}^{G}{\entropy{X_g}} \right) - \entropy{X_1,X_2,\ldots,X_G}
	\label{eq:multiInfo}.
\end{align}
Equivalently we can split the set into two parts, $\mathbf{X} = \left\{ \mathbf{Y}, \mathbf{Z}\right\}$, and express this quantity iteratively in terms of the multi-information of its components individually and the mutual information between those components:
\begin{align}
	\multiInfo{\mathbf{X}} = \multiInfo{\mathbf{Y}} + \multiInfo{\mathbf{Z}} + \mutualInfo{\mathbf{Y}}{\mathbf{Z}}
	\label{eq:multiInfoIterative}.
\end{align}

The \textbf{conditional mutual information} between $X$ and $Y$ given $Z$ is the mutual information between $X$ and $Y$ when $Z$ is known:
\begin{align}
	\condMutualInfo{X}{Y}{Z} & = \sum_{x \in \alpha_x}{\sum_{y \in \alpha_y}{\sum_{z \in \alpha_z}{ p(x,y,z) \log_2{\frac{p(x \mid y,z)}{p(x \mid z)}} }}}
	\label{eq:condMiDirectional} \\
							& = \sum_{x \in \alpha_x}{\sum_{y \in \alpha_y}{\sum_{z \in \alpha_z}{ p(x,y,z) \log_2{\frac{p(x,y,z)p(z)}{p(x, z)p(y, z)}} }}}
	\label{eq:condMiNonDirectional} \\
		& = \condEntropy{X}{Z} - \condEntropy{X}{Y,Z}
	\label{eq:condMi}.
\end{align}
Note that a conditional MI $\condMutualInfo{X}{Y}{Z}$ may be either larger or smaller than the related unconditioned MI $I(X;Y)$ \citep{mac03}.
Such conditioning removes redundant information in $Y$ and $Z$ about $X$, but adds synergistic information which can only be decoded with knowledge of both $Y$ and $Z$ (see further description regarding ``partial information decomposition'', which refers to attempts to tease these components apart, by:  \citep{will10a,hard12a,grif14a,liz13b,bert12a}).

One can consider the MI from two variables $Y_1, Y_2$ jointly to another variable $X$, $I(X; Y_1,Y_2)$, and using \eq{entropyChainRule}, \eq{mi} and \eq{condMi} decompose this into the information carried by the first variable plus that carried by the second conditioned on the first:
\begin{align}
	\mutualInfo{X}{Y_1,Y_2} = \mutualInfo{X}{Y_1} + \condMutualInfo{X}{Y_2}{Y_1}
	\label{eq:chainRuleMi}.
\end{align}
Of course, this \emph{chain rule} generalises to multivariate $\mathbf{Y}$ of dimension greater than two.

\subsection{Measures of information dynamics}
\label{app:infoDynamicsMeasures}

Next, we build on the basic measures of information theory to present measures of the dynamics of information processing.
We focus on measures of information in \emph{time-series processes} $X$ of the random variables $\{ \ldots X_{n-1}, X_{n}, X_{n+1} \ldots \}$ with process realisations $\{ \ldots x_{n-1}, x_{n}, x_{n+1} \ldots \}$ for countable time indices $n$.

We briefly review the framework for \emph{information dynamics} which was recently introduced by \citet{liz07b,liz08a,liz10e,liz12a,liz14a} and \citet{liz13a,liz14b}.
This framework considers how the information in variable $X_{n+1}$ is related to previous variables, e.g. $X_{n}$, of the process or other processes, addressing the fundamental question: \emph{``where does the information in a random variable $X_{n+1}$ in a time series come from?''}.
As indicated in \fig{infoDynamics}, this question is addressed in terms of information from the past of process $X$ (i.e. the information \emph{storage}), information contributed from other source processes $Y$ (i.e. the information \emph{transfer}), and how these sources combine (information \emph{modification}).
The goal is to decompose the information in the next observation of $X$, $X_{n+1}$, in terms of these information sources.

The \textbf{entropy rate} is defined as \citep{cover91}:
\begin{align}
	\entropyRate{X} & = \lim_{n \rightarrow \infty}{\frac{1}{n} \entropy{X_1,X_2,\ldots, X_n}} \\
			& = \lim_{n \rightarrow \infty}{\frac{1}{n} \entropy{ \mathbf{X}^{(n)}_n }}
	\label{eq:entropyRate},
\end{align}
(where the limit exists) where we have used $\mathbf{X}^{(k)}_n = \left\{ X_{n-k+1}, \ldots , X_{n-1}, X_n \right\}$ to denote the $k$ consecutive variables of $X$ up to and including time step \textit{n}, which has realizations $\mathbf{x}^{(k)}_n = \left\{ x_{n-k+1}, \ldots , x_{n-1}, x_n \right\}$. 
This quantity describes the limiting rate at which the entropy of $n$ consecutive measurements of $X$ grow with $n$.
A related definition for a \textbf{(conditional) entropy rate} is given by:\footnote{Note that we have reversed the use of the primes in the notation from \citet{cover91}, in line with \citet{crutch03}.}
\begin{align}
	\entropyRateCond{X} & = \lim_{n \rightarrow \infty}{ \condEntropy{X_n}{X_1,X_2,\ldots, X_{n-1}}} \\
			& = \lim_{n \rightarrow \infty}{ \condEntropy{X_n}{\mathbf{X}_{n-1}^{(n-1)}} }
	\label{eq:entropyRateConditional}.
\end{align}
For stationary processes $X$, the limits for the two quantities $\entropyRate{X}$ and $\entropyRateCond{X}$ exist (i.e. the expected entropy rate converges) and are equal \citep{cover91}.

For our purposes in considering information dynamics, we are interested in the conditional formulation $\entropyRateCond{X}$, since it explicitly describes how one random variable $X_n$ is related to the previous instances $\mathbf{X}_{n-1}^{(n-1)}$.
For practical usage, we are particularly interested in estimation of $\entropyRateCond{X}$ with finite-lengths $k$, and in estimating it regarding the information at different time indices $n$.
That is to say, we use the notation $\entropyRateCondK{X_{n+1}}{k}$ to describe finite-$k$ estimates of the conditional entropy rate in $X_{n+1}$ given $\mathbf{X}_{n}^{(k)}$:
\begin{align}
	\entropyRateCondK{X_{n+1}}{k} = \condEntropy{X_{n+1}}{\mathbf{X}_{n}^{(k)}}
	\label{eq:entropyRateConditionalK}.
\end{align}
Assuming stationarity we define:
\begin{align}
	\entropyRateCondK{X}{k} = \entropyRateCondK{X_{n+1}}{k}
	\label{eq:entropyRateConditionalKSTat}.
\end{align}
for any $n$, and of course letting $k = n$ and joining \eq{entropyRateConditional} and \eq{entropyRateConditionalK} we have $\lim_{n \rightarrow \infty} \entropyRateCondK{X_{n+1}}{k} = \entropyRateCond{X}$.

Next, the \textbf{effective measure complexity} \citep{grass86a} or \textbf{excess entropy} \citep{crutch03} quantifies the total amount of structure or memory in the process $X$, and is computed in terms of the slowness of the approach of the conditional entropy rate estimates to their limiting value:
\begin{align}
	\excessEntropy{X} = \sum_{k=0}^{\infty}{\left( \entropyRateCondK{X}{k} - \entropyRateCond{X} \right)}
	\label{eq:excessEntropy}.
\end{align}
When the process $X$ is stationary 
we may represent the excess entropy as the mutual information between the semi-infinite past and semi-infinite future of the process:
\begin{align}
	\excessEntropy{X} & = \lim_{k \rightarrow \infty}{\excessEntropyK{X}{k}}
	\label{eq:predictiveInfo}, \\
	\excessEntropyK{X}{k} & = {\mutualInfo{\mathbf{X}_n^{(k)}}{\mathbf{X}_{n+1}^{(k^+)}}}
	\label{eq:predictiveInfoFiniteK},
\end{align}
where $\mathbf{X}_{n+1}^{(k^+)}$ refers to the next $k$ values $\left\{ X_{n+1}, X_{n+2}, \ldots , X_{n+k} \right\}$ with realizations $\mathbf{x}_{n+1}^{(k^+)} = \left\{ x_{n+1}, x_{n+2}, \ldots , x_{n+k} \right\}$, and $\excessEntropyK{X}{k}$ are finite-$k$ estimates of $\excessEntropy{X}$.
This formulation is known as the \textbf{predictive information} \citep{bialek01}, as it highlights that the excess entropy captures the information in a system's past which can also be found in its future.
It is the most appropriate formulation for our purposes, since it provides a clear interpretation as information storage.
That is, the excess entropy can be viewed in this formulation as measuring information from the past of the process that is stored -- potentially in a distributed fashion in external variables -- and is used at some point in the future of the process \citep{liz12a}.
This contrasts with the statistical complexity \citep{crutch89,sha01a}, an upper bound to the excess entropy, which measures \textit{all} information which is \textit{relevant} to the prediction of the future of the process states; i.e. the stored information which \textit{may be used} in the future \citep{liz12a}.

In contrast again, the \textbf{active information storage} (AIS) was introduced by \citet{liz12a} to measure how much of the information from the past of the process $X$ is observed to be \emph{in use} in computing its \emph{next observation}.
This measure of information storage more directly addresses our key question of determining the sources of the information in the next observation $X_{n+1}$.
The active information storage is the expected mutual information between realizations $\mathbf{x}^{(k)}_{n}$ of the past state $\mathbf{X}_n^{(k)}$ (as $k \rightarrow \infty$) and the corresponding realizations $x_{n+1}$ of the next value $X_{n+1}$ of process $X$:
\begin{align}
	\ais{X} & = \lim_{k \rightarrow \infty}{\aisK{X}{k}}
		\label{eq:activeStorage}, \\
	\aisK{X}{k} & = \mutualInfo{\mathbf{X}_n^{(k)}}{X_{n+1}}
		\label{eq:activeStorageEstimate}.
\end{align}
We note that $\mathbf{x}_n^{(k)}$ are Takens' \textit{embedding vectors} \citep{takens81} with \textit{embedding dimension} $k$, which capture the underlying \textit{state} of the process $X$ for Markov processes of order $k$.\footnote{We can use an embedding delay $\tau$ to give $\mathbf{x}_n^{(k)} = \left\{ x_{n-(k-1)\tau}, \ldots , x_{n-\tau}, x_n \right\}$, where this helps to better empirically capture the state from a finite sample size. Non-uniform embeddings (i.e. with irregular delays) may also be useful \citep{faes11a} (not implemented in JIDT at this stage).}
As such, one needs to at least take $k$ at the Markovian order of $X$ in order to capture all relevant information in the past of $X$, otherwise (for non-Markovian processes) the limit $k \rightarrow \infty$ is theoretically required in general \citep{liz12a}.
We also note that since:
\begin{align}
	\ais{X} = \entropy{X} - \entropyRateCond{X}
	\label{eq:relatingActiveToHAndHmu},
\end{align}
then the limit in \eq{activeStorage} exists for stationary processes (i.e. $A(X)$ converges with $k \rightarrow \infty$) \citep{liz12a}.

Arguably the most important measure in this toolkit is the \textbf{transfer entropy} (TE) from \citet{schr00}.
TE captures the concept of information transfer, as the amount of information that a source process provides about a destination (or target) process' next state in the context of the destination's past.
Quantitatively, this is the expected mutual information from realizations $\mathbf{y}^{(l)}_{n}$ of the state $\mathbf{Y}_n^{(l)}$ of a source process $Y$ to the corresponding realizations $x_{n+1}$ of the next value $X_{n+1}$ of the destination process $X$, conditioned on realizations $\mathbf{x}^{(k)}_{n}$ of its previous state $\mathbf{X}_n^{(k)}$:
\begin{align}
	\teArgs{Y}{X}{l} & = \lim_{k \rightarrow \infty}{\teArgs{Y}{X}{k,l}}
	\label{eq:te}, \\
	\teArgs{Y}{X}{k,l} & = \condMutualInfo{\mathbf{Y}_n^{(l)}}{X_{n+1}}{\mathbf{X}_n^{(k)}}
	\label{eq:teLimit}.
\end{align}
TE has become a very popular tool in complex systems in general, e.g. \citep{will11a,lung06,obs10a,barn12a,liz08a,liz11b,boed12a}, and in computational neuroscience in particular, e.g. \citep{vic11a,lind11a,ito11a,stra12a,liz11a}.
For multivariate Gaussians, the TE is equivalent (up to a factor of 2) to the \textbf{Granger causality} \citep{barn09}.

There are a number of important considerations regarding the use of this measure (see further discussion by \citet{liz08a}; \citet{liz14b}; \citet{wib14b,wib14c}; and \citet{vic14a}).
First, for the embedding vectors $\mathbf{x}_n^{(k)}$ one needs to at least take $k$ larger than the Markovian order of $X$ in order to eliminate any AIS from being redundantly measured in the TE.\footnote{The destination's embedding dimension should be increased before that of the source, for this same reason.}
Then, one may need to extend $k$ to capture synergies generated in $x_{n+1}$ between the source $\mathbf{y}_n^{(l)}$ and earlier values in $X$.
For non-Markovian processes $X$ (or non-Markovian processes when considered jointly with the source), one should theoretically take the limit as $k \rightarrow \infty$ \citep{liz08a}.
Setting $k$ in this manner gives the perspective to separate information storage and transfer in the distributed computation in process $X$, and allows one to interpret the transfer entropy as properly representing information transfer \citep{liz08a,liz10a}.

Also, note that the transfer entropy can be defined for an arbitrary source-destination delay $u$ \citep{wib13a}:
\begin{align}
	\teArgs{Y}{X}{k,l,u} & = \condMutualInfo{\mathbf{Y}_{n+1-u}^{(l)}}{X_{n+1}}{\mathbf{X}_n^{(k)}}
	\label{eq:teDelay},
\end{align}
and indeed that this should be done for the appropriate causal delay $u>0$ .
For ease of presentation here, we describe the measures for $u=1$ only, though all are straightforward to generalise and are implemented with generic $u$ in JIDT.

While the simple setting $l=1$ is often used, this is only completely appropriate where $y_n$ is directly causal to $x_{n+1}$ and where it is the only direct causal source in $Y$ \citep{liz08a,liz10a} (e.g. in cellular automata).
In general circumstances, one should use an embedded source \emph{state} $\mathbf{y}^{(l)}_{n}$ (with $l>1$), in particular where the observations $y$ mask a hidden Markov process that is causal to $X$ (e.g. in brain imaging data), or where multiple past values of $Y$ in addition to $y_n$ are causal to $x_{n+1}$.

Finally, for proper interpretation as information transfer, $Y$ is constrained among the causal information contributors to $X$ \citep{liz10a}. With that said, the concepts of information transfer and causality are complementary but distinct, and TE should not be thought of as measuring causal effect \citep{ay08,liz10a,chich12a}.
\citet{pro13a} and \citet{pro14b} have also provided a thermodynamic interpretation of transfer entropy, as being proportional to external entropy production, possibly due to irreversibility.

Now, the transfer entropy may also be conditioned on other possible sources $Z$ to account for their effects on the destination.
The \textbf{conditional transfer entropy}\footnote{This is sometimes known as ``multivariate'' TE, though this term can be confused with TE applied to multivariate source and destination variables (i.e. the collective TE).} was introduced for this purpose \citep{liz08a,liz10e}:
\begin{align}
	\condTeArgs{Y}{X}{Z}{l} & = \lim_{k \rightarrow \infty}{\condTeArgs{Y}{X}{Z}{k,l}}
	\label{eq:teCondLimit}, \\
	\condTeArgs{Y}{X}{Z}{k,l} & = \condMutualInfo{\mathbf{Y}_n^{(l)}}{X_{n+1}}{\mathbf{X}_n^{(k)},Z_n}
	\label{eq:teCondEstimate},
\end{align}
Note that $Z_n$ may represent an embedded state of another variable, or be explicitly multivariate.
Also, for simplicity \eq{teCondEstimate} does not explicitly show arbitrary delays in the style of \eq{teDelay} for source-destination and conditional-destination relationships, though these may naturally be defined and are implemented in JIDT.
Transfer entropies conditioned on other variables have been used in several biophysical and neuroscience applications, e.g. \citep{faes11a,faes12a,stra12a,vak09}.
We typically describe TE measurements which are not conditioned on any other variables (as in \eq{te}) as \textbf{pairwise} or \textbf{apparent transfer entropy}, and measurements conditioned on \emph{all} other causal contributors to $X_{n+1}$ as \textbf{complete transfer entropy} \citep{liz08a}.
Further, one can consider multivariate sources $\mathbf{Y}$, in which case we refer to the measure $\teArgs{\mathbf{Y}}{X}{k,l}$ as a \textbf{collective transfer entropy} \citep{liz10e}.

Finally, while how to measure information modification remains an open problem (see \citet{liz13b}), JIDT contains an implementation of an early attempt at capturing this concept in the \textbf{separable information} \citep{liz10e}:
\begin{align}
	\separable{X} & = \lim_{k \rightarrow \infty}{ \separableK{X}{k} }
		\label{eq:separableInfo}, \\
	\separableK{X}{k} & = \aisK{X}{k} + \sum_{Y \in \mathbf{V}_X \setminus X}{ \teArgs{Y}{X}{k,l_Y} }
		\label{eq:separableInfoEstimate}.
\end{align}
Here, $\mathbf{V}_X$ represents the set of causal information sources $\mathbf{V}_X$ to $X$, while $l_Y$ is the embedding dimension for source $Y$.

\subsection{Local information-theoretic measures}
\label{app:localMeasures}

\textbf{Local information-theoretic measures} (also known as \textbf{pointwise information-theoretic measures}) characterise the information attributed with \emph{specific} measurements $x$, $y$ and $z$ of variables $X$, $Y$ and $Z$ \citep{liz14b}, rather than the traditional expected or average information measures associated with these variables introduced in \appRefSec{classicMeasures} and \appRefSec{infoDynamicsMeasures}.
Although they are deeply ingrained in the fabric of information theory, and heavily used in some areas (e.g. in natural language processing \citep{man99}), until recently \citep{sha01a,sha06,hel04,liz07b,liz08a,liz12a,liz10e} local information-theoretic measures were rarely applied to complex systems.

That these local measures are now being applied to complex systems is important, because they provide a direct, model-free, mechanism to analyse the \emph{dynamics} of how information processing unfolds in time.
In other words: traditional (expected) information-theoretic measures would return one value to characterise, for example, the transfer entropy between $Y$ and $X$.
Local transfer entropy on the other hand, returns a time-series of values to characterise the information transfer from $Y$ to $X$ as a function of time, so as to directly reveal the \emph{dynamics} of their interaction.
Indeed, it is well-known that local values (within a global average) provide important insights into the dynamics of nonlinear systems \citep{das02}.

A more complete description of local information-theoretic measurements is provided by \citet{liz14b}.
Here we provide a brief overview of the local values of the measures previously introduced.

The most illustrative local measure is of course the \textbf{local entropy} or \textbf{Shannon information content}.
The Shannon information content of an outcome $x$ of measurement of the variable $X$ is \citep{mac03}: 
\begin{align}
	\localEntropy{x} = - \log_2{p(x)}
	\label{eq:localEntropy}.
\end{align}
Note that by convention we use lower-case symbols to denote local information-theoretic measures.
The Shannon information content was shown to be the unique formulation for a local entropy (up to the base of the logarithm) satisfying required properties corresponding to those of the expected Shannon entropy (see \citet{ash65} for details).
Now, the quantity $\localEntropy{x}$ is simply the information content attributed to the specific symbol $x$, or the information required to predict or uniquely specify that specific value.
Less probable outcomes $x$ have higher information content than more probable outcomes, and we have $\localEntropy{x} \geq 0$.
The Shannon information content of a given symbol $x$ is the \emph{code-length} for that symbol in an optimal encoding scheme for the measurements $X$, i.e. one that produces the minimal expected code length.

We can form all traditional information-theoretic measures as the \emph{average} or \emph{expectation value} of their corresponding local measure, e.g.:
\begin{align}
	\entropy{X} & = \sum_{x \in \alpha_x}{p(x) \localEntropy{x}}
	\label{eq:localEntropyAveraged}, \\
	& = \left\langle h(x) \right\rangle
	\label{eq:localEntropyExpectation}.
\end{align}
While the above represents this as an expectation over the relevant ensemble, we can write the same average over all of the $N$ samples $x_n$ (with each sample given an index $n$) used to generate the probability distribution function (PDF) $p(x)$ \citep{liz14b,liz08a}, e.g.:
\begin{align}
	\entropy{X} & = \frac{1}{N} \sum_{n=1}^{N}{\localEntropy{x_n}}
	\label{eq:localEntropyAveragedN}, \\
	& = \left\langle \localEntropy{x_n} \right\rangle_n
	\label{eq:localEntropyExpectationN}.
\end{align}

Next, we have the \textbf{conditional Shannon information content} (or \textbf{local conditional entropy}) \citep{mac03}: 
\begin{align}
	\localCondEntropy{x}{y} & = -\log_2{p(x \mid y)}
	\label{eq:localConditional}, \\
	\localJointEntropy{x}{y} & = -\log_2{p(x, y)}
	\label{eq:localJointEntropy}, \\
	& = \localEntropy{y} + \localCondEntropy{x}{y}
	\label{eq:localConditionalChainRule}, \\
	\condEntropy{X}{Y} & = \left\langle \localCondEntropy{x}{y} \right\rangle
	\label{eq:localConditionalAveraged}.
\end{align}
As above, local quantities satisfy corresponding chain rules to those of their expected quantities.

The \textbf{local mutual information} is defined (uniquely, see \citet[ch. 2]{fano61}) as ``the amount of information provided by the occurrence of the event represented by $y_i$ about the occurrence of the event represented by $x_i$'', i.e.:
\begin{align}
	\localMutualInfo{x}{y} & = \log_2{\frac{p(x \mid y)}{p(x)}}
	\label{eq:localMi}, \\
		& = \localEntropy{x} - \localCondEntropy{x}{y}
	\label{eq:localMiFromShannon}, \\
	\mutualInfo{X}{Y} & = \left\langle \localMutualInfo{x}{y} \right\rangle
	\label{eq:localMiAveraged}.
\end{align}
$\localMutualInfo{x}{y}$ is symmetric in $x$ and $y$, as is the case for $\mutualInfo{x}{y}$.
The local mutual information is the difference in code lengths between coding the value $x$ in isolation (under the optimal encoding scheme for $X$), or coding the value $x$ given $y$ (under the optimal encoding scheme for $X$ given $Y$).
In other words, this quantity captures the coding ``cost'' for $x$ in not being aware of the value $y$.

Of course this ``cost'' averages to be non-negative, however the local mutual information may be either positive or negative for a specific pair $x,y$.
Positive values are fairly intuitive to understand: $\localMutualInfo{x}{y}$ is positive where $p(x \mid y) > p(x)$, i.e. knowing the value of $y$ \textit{increased} our expectation of (or positively informed us about) the value of the measurement $x$.
Negative values simply occur in \eq{localMi} where $p(x \mid y) < p(x)$.
That is, knowing the value of $y$ changed our belief $p(x)$ about the probability of occurrence of the outcome $x$ to a smaller value $p(x \mid y)$, and hence we considered it less likely that $x$ would occur when knowing $y$ than when not knowing $y$, in a case were $x$ nevertheless occurred. 
Consider the following example from \citet{liz14b}, of the probability that it will rain today, $p(\texttt{rain}=1)$, and the probability that it will rain given that the weather forecast said it would not, $p(\texttt{rain}=1 \mid \texttt{rain\_forecast}=0)$.
We could have $p(\texttt{rain}=1 \mid \texttt{rain\_forecast}=0) < p(\texttt{rain}=1)$, so we would have $i(\texttt{rain}=1 ; \texttt{rain\_forecast}=0) < 0$, because we considered it less likely that rain would occur today when hearing the forecast than without the forecast, in a case where rain nevertheless occurred.
Such negative values of MI are actually quite meaningful, and can be interpreted as there being negative information in the value of $y$ about $x$. We could also interpret the value $y$ as being \textit{misleading} or \textit{misinformative} about the value of $x$, because it \textit{lowered} our expectation of observing $x$ prior to that observation being made in this instance. In the above example, the weather forecast was misinformative about the rain today.

Note that the local mutual information $\localMutualInfo{x}{y}$ measure above is distinct from \emph{partial} localization expressions, i.e. the partial mutual information or specific information $\mutualInfo{x}{Y}$ \citep{dewe99}, which consider information contained in specific values $x$ of one variable $X$ about the other (unknown) variable $Y$.
While there are two valid approaches to measuring partial mutual information, as above there is only one valid approach for the fully local mutual information $\localMutualInfo{x}{y}$ \citep[ch. 2]{fano61}.

The \textbf{local conditional mutual information} is similarly defined by \citet[ch. 2]{fano61}:
\begin{align}
	\localCondMutualInfo{x}{y}{z} & = \log_2{\frac{p(x \mid y,z)}{p(x \mid z)}}
	\label{eq:localCondMi}, \\
		& = \localCondEntropy{x}{z} - \localCondEntropy{x}{y,z}
	\label{eq:localCondMiFromShannon}, \\
	\condMutualInfo{X}{Y}{Z} & = \left\langle \localCondMutualInfo{x}{y}{z} \right\rangle
	\label{eq:localCondMiAveraged}.
\end{align}
$\condMutualInfo{X}{Y}{Z}$ is the difference in code lengths (or coding cost) between coding the value $x$ given $z$ (under the optimal encoding scheme for $X$ given $Z$), or coding the value $x$ given both $y$ and $z$ (under the optimal encoding scheme for $X$ given $Y$ and $Z$).
As per $\mutualInfo{X}{Y}$, the local conditional MI is symmetric in $x$ and $y$, and may take positive or negative values.

The \textbf{local multi-information} follows for the observations $x_1,x_2, \ldots ,x_G$ as:
\begin{align}
\localMultiInfoPlus{x_1}{x_2}{x_G} = \left( \sum_{g=1}^{G}{\localEntropy{x_g}} \right) - \localEntropy{x_1,x_2,\ldots,x_G}
\label{eq:localMultiInfo}.
\end{align}

Local measures of information dynamics are formed via the local definitions of the basic information-theoretic measures above.
Here, the local measures pertain to realisations $x_n$, $\mathbf{x}_{n}^{(k)}$, $\mathbf{y}_{n}^{(l)}$, etc, of the processes at specific time index $n$.
The PDFs may be estimated either from multiple realisations of the process for time index $n$, or from multiple observations over time from one (or several) full time-series realisation(s) where the process is stationary (see comments by \citet{liz14b}).

We have the \textbf{local entropy rate}:\footnote{For the local measures of information dynamics, while formal definitions may be provided by taking the limit as $k \rightarrow \infty$, we will state only the formulae for their finite-$k$ estimates.}
\begin{align}
	\localEntropyRateCond{X}{n+1,k} & = \localCondEntropy{x_{n+1}}{\mathbf{x}_{n}^{(k)}}
	\label{eq:localEntropyRateConditionalK}, \\
	\entropyRateCondK{X}{k} & = \left\langle \localEntropyRateCond{X}{n,k} \right\rangle
	\label{eq:localEntropyRateConditionalKAveraged}.
\end{align}

Next, the \textbf{local excess entropy} is defined as (via the predictive information formulation from \eq{predictiveInfoFiniteK}) \citep{sha01a}:
\begin{align}
	\localExcessEntropy{X}{n+1,k} & = \localMutualInfo{\mathbf{x}_n^{(k)}}{\mathbf{x}_{n+1}^{(k^+)}}
	\label{eq:localPredictiveInfoFiniteK}, \\
	\excessEntropyK{X}{k} & = \left\langle \localExcessEntropy{X}{n,k} \right\rangle
	\label{eq:localPredictiveInfoFiniteKAveraged}.
\end{align}

We then have the \textbf{local active information storage} $\localAis{X}{n+1}$ \citep{liz12a}:
\begin{align}
	\localAis{X}{n+1,k} & = \localMutualInfo{\mathbf{x}^{(k)}_n}{x_{n+1}}
		\label{eq:localActiveK}, \\
	\aisK{X}{k} & = \left\langle \localAis{X}{n+1,k} \right\rangle
		\label{eq:localActiveKAveraged}.
\end{align}
The local values of active information storage measure the dynamics of information storage at different time points within a system, revealing to us how the use of memory fluctuates during a process.
As described for the local MI, $\localAis{X}{n+1,k}$ may be positive or negative, meaning the past history of the process can either positively inform us or actually \emph{misinform} us about its next value \citep{liz12a}.
\fig{infoDynamics} indicates a local active information storage measurement for time-series process $X$.

The \textbf{local transfer entropy} is \citep{liz08a} (with adjustment for source-destination lag $u$ \citep{wib13a}):
\begin{align}
	\localTe{Y}{X}{n+1, k, l, u} & = \localCondMutualInfo{\mathbf{y}^{(l)}_{n+1-u}}{x_{n+1}}{\mathbf{x}^{(k)}_{n}}
		\label{eq:localTEK}, \\
	\te{Y}{X}{k,l} & = \left\langle\localTe{Y}{X}{n+1, k, l} \right\rangle
		\label{eq:localTeKAveraged}.
\end{align}
These local information transfer values measure the dynamics of transfer in time between a given pair of time-series processes, revealing to us how information is transferred in time and space.
\fig{infoDynamics} indicates a local transfer entropy measurement for a pair of processes $Y \rightarrow X$.

Finally, we have the \textbf{local conditional transfer entropy} \citep{liz08a,liz10e} (again dropping arbitrary lags and embedding of the conditional here for convenience):
\begin{align}
	\localCondTe{Y}{X}{Z}{n+1, k, l} & = \localCondMutualInfo{\mathbf{y}^{(l)}_{n}}{x_{n+1}}{\mathbf{x}^{(k)}_{n}, z_n}
		\label{eq:localCondTEK}, \\
	\condTeK{Y}{X}{Z}{n+1, k, l} & = \left\langle \localCondTe{Y}{X}{Z}{n+1, k, l} \right\rangle
		\label{eq:localCondTEKAveraged}.
\end{align}

\subsection{Differential entropy}
\label{app:differentialEntropy}

Note that all of the information-theoretic measures above considered a discrete alphabet of symbols $\alpha_x$ for a given variable $X$.
When $X$ in fact is a continuous-valued variable, we shift to consider \textbf{differential entropy} measurements; see \citet[ch. 9]{cover91}.
We briefly discuss differential entropy, since some of our estimators discussed in \appRefSec{estimationTechniques} evaluate these quantities for continuous-valued variables rather than strictly Shannon entropies.

The differential entropy of a continuous variable $X$ with probability density function $f(x)$ is defined as \cite[ch. 9]{cover91}:
\begin{align}
	\diffEntropy{X} = - \int_{S_X} {f(x) \log{f(x)}} dx
	\label{eq:diffEntropy},
\end{align}
where $S_X$ is the set where $f(x) > 0$.
The differential entropy is strongly related to the Shannon entropy, but has important differences to what the Shannon entropy would return on discretizing the same variables.
Primary amongst these differences is that $\diffEntropy{X}$ changes with scaling of the variable $X$, and that it can be negative.

Joint and conditional ($\condDiffEntropy{X}{Y}$) differential entropies may be evaluated from \eq{diffEntropy} expressions using the same chain rules from the Shannon measures.
Similarly, the differential mutual information may be defined as \cite[ch. 9]{cover91}:
\begin{align}
	\diffMI{X}{Y} & = \diffEntropy{X} - \condDiffEntropy{X}{Y}
	\label{eq:diffMiChainRule}, \\
		& = \int_{S_X,S_Y} {f(x,y) \log{\frac{f(x,y)}{f(x)f(y)}}} \ dx \ dy
	\label{eq:diffMi}.
\end{align}
Crucially, the properties of $\diffMI{X}{Y}$ are the same as for discrete variables, and indeed $\diffMI{X}{Y}$ is equal to the discrete MI $\mutualInfo{X^\Delta}{Y^\Delta}$ for discretizations $X^\Delta$ and $Y^\Delta$ with bin size $\Delta$, in the limit $\Delta \rightarrow 0$ \cite[ch. 9]{cover91}.
Conditional MI and other derived measures (e.g. transfer entropy) follow.

\subsection{Statistical significance testing}
\label{app:statSigTesting}

In \textit{theory}, the MI between two unrelated variables $Y$ and $X$ is equal to 0.
The same goes for the TE between two variables $Y$ and $X$ with no directed relationship, or the conditional MI between $Y$ and $X$ given $Z$ where there is no conditional relationship.
In \textit{practice}, where the MI, conditional MI or TE are empirically measured from a finite number of samples $N$, a bias of a non-zero measurement is likely to result even where there is no such (directed) relationship.
A common question is then whether a given empirical measurement is statistically different from 0, and therefore represents sufficient evidence for a (directed) relationship between the variables.

This question is addressed in the following manner \citep{cha03,verd05,vic11a,lind11a,liz11a,wib14c,barn12a}.
We form a \textit{null hypothesis} $H_0$ that there is no such relationship, and then make a test of statistical significance of evidence (our original measurement) in support of that hypothesis.
To perform such a test, we need to know what the \emph{distribution} for our measurement would look like if $H_0$ was true, and then evaluate a $p$-value for sampling our actual measurement from this distribution.
If the test fails, we accept the alternate hypothesis that there is a (directed) relationship.

For example, for an MI measurement $\mutualInfo{Y}{X}$, we generate the distribution of \emph{surrogate} measurements $\mutualInfo{Y^s}{X}$ under the assumption of $H_0$.
Here, $Y^s$ represents \emph{surrogate} variables for $Y$ generated under $H_0$, which have the same statistical properties as $Y$, but any potential correlation with $X$ is destroyed.
Specifically, this means that $p(x \mid y)$ in \eq{mi} is distributed as $p(x)$ (with $p(y)$ retained also).

In some situations, we can compute the distribution of $\mutualInfo{Y^s}{X}$ analytically.
For example, for linearly-coupled Gaussian multivariates $\mathbf{X}$ and $\mathbf{Y}$, $\mutualInfo{\mathbf{Y}^s}{\mathbf{X}}$ measured in \textit{nats} follows a chi-square distribution, specifically $\chi^2_{|\mathbf{X}| |\mathbf{Y}|}/2N$ with $|\mathbf{X}| |\mathbf{Y}|$ degrees of freedom, where $|\mathbf{X}|$ ($|\mathbf{Y}|$) is the number of Gaussian variables in vector $\mathbf{X}$ ($\mathbf{Y}$) \citep{gew82,bri04}.
Also, for discrete variables $X$ and $Y$ with alphabet sizes $M_X$ and $M_Y$, $\mutualInfo{Y^s}{X}$ measured in \textit{bits} follows a chi-square distribution, specifically $\chi^2_{(M_X-1)(M_Y-1)}/(2N \log{2})$ \citep{bri04,cheng06}.
Note that these distributions are followed \textit{asymptotically} with the number of samples $N$, and the approach is much slower for discrete variables with skewed distributions \citep{barn13PersComm}, \textit{which reduces the utility of this analytic result in practice}.\footnote{See \secRef{otherDemos} for an investigation of this.}
\citet{barn12a} generalise these results to state that a model-based null distribution (in \textit{nats}) will follow $\chi^2_{d}/2N$, where $d$ is the ``difference between the number of parameters'' in a full model (capturing $p(x \mid y)$ in \eq{mi}) and a null model (capturing $p(x)$ only).

Where no analytic distribution is known, the distribution of $\mutualInfo{Y^s}{X}$ must be computed empirically.
This is done by a resampling method (i.e. permutation or bootstrapping)\footnote{JIDT employs permutation tests for resampling.} \citep{cha03,verd05,vic11a,lind11a,liz11a,wib14c}, creating a large number of surrogate time-series pairs $\{Y^s,X\}$ by shuffling (for permutations, or redrawing for bootstrapping) the samples of $Y$ (so as to retain $p(x)$ and $p(y)$ but not $p(x \mid y)$), and computing a population of $\mutualInfo{Y^s}{X}$ values.

Now, for a conditional MI, we generate the distribution of $\condMutualInfo{Y^s}{X}{Z}$ under $H_0$, which means that $p(x \mid y, z)$ in \eq{condMiDirectional} is distributed as $p(x \mid z)$ (with $p(y)$ retained also).\footnote{Clearly, this approach specifically makes a \emph{directional} hypothesis test of \eq{condMiDirectional} rather than a non-directional test of \eq{condMiNonDirectional}. Asymptotically these will be the same anyway (as is clear for the analytic cases discussed here). In practice, we favour this somewhat directional approach since in most cases we are indeed interested in the directional question of whether $Y$ adds information to $X$ in the context of $Z$.}
The asymptotic distribution may be formed analytically for linearly-coupled Gaussian multivariates defined above \citep{gew82,barn12a}
(in \textit{nats}) as $\chi^2_{|\mathbf{X}| |\mathbf{Y}|}/2N$ with $|\mathbf{X}| |\mathbf{Y}|$ degrees of freedom -- interestingly, this does \textit{not} depend on the $Z$ variable.
Similarly, for discrete variables the asymptotic distribution (in \textit{bits}) is $\chi^2_{(M_X-1)(M_Y-1)M_Z}/(2N \log{2})$ \citep{cheng06}.
Again, the distribution of $\condMutualInfo{Y^s}{X}{Z}$ is otherwise computed by permutation (or bootstrapping), this time by creating surrogate time-series $\{Y^s,X,Z\}$ by shuffling (or redrawing) the samples of $Y$ (retaining $p(x \mid z)$ and $p(y)$ but not $p(x \mid y, z)$), and computing a population of $\condMutualInfo{Y^s}{X}{Z}$ values.

Statistical significance testing for the transfer entropy can be handled as a special case of the conditional MI.
For linear-coupled Gaussian multivariates $\mathbf{X}$ and $\mathbf{Y}$, the null $\teArgs{Y^s}{X}{k,l}$ (in \textit{nats}) is asymptotically $\chi^2/2N$ distributed with $l |\mathbf{X}| |\mathbf{Y}|$ degrees of freedom \citep{gew82,barn12a,barn13PersComm}, while for discrete $X$ and $Y$, $\teArgs{Y^s}{X}{k,l}$ (in \textit{bits}) is asymptotically $\chi^2/(2N \log{2})$ distributed with $(M_X-1)(M_Y^l-1)M_X^k$ degrees of freedom \citep{barn12a}.
Again, the distribution of $\teArgs{Y^s}{X}{k,l}$ is otherwise computed by permutation (or bootstrapping) \citep{cha03,verd05,vic11a,lind11a,liz11a,wib14c}, under which surrogates must preserve $p(x_{n+1} \mid x^{(k)}_n)$ but not $p(x_{n+1} \mid \mathbf{x}^{(k)}_n, \mathbf{y}^{(l)}_n)$.
Directly shuffling the series $Y$ to create the $Y^s$ is \textit{not} a valid approach, since it destroys $\mathbf{y}^{(l)}_n$ vectors (unless $l=1$).
Valid approaches include: shuffling (or redrawing) the $\mathbf{y}^{(l)}_n$ amongst the set of $\{x_{n+1}, \mathbf{x}^{(k)}_n, \mathbf{y}^{(l)}_n \}$ tuples;\footnote{This is the approach taken in JIDT.} rotating the $Y$ time-series (where we have stationarity); or swapping sample time series $Y_i$ between different trials $i$ in an ensemble approach \citep{vic11a,wib14c,lind11a,woll14a}.
Conditional TE may be handled similarly as a special case of a conditional MI.

Finally, we note that such assessment of statistical significance is often used in the application of effective network inference from multivariate time-series data; e.g. \citep{vic11a,lind11a,liz11a,wib14c}.
In this and other situations where multiple hypothesis tests are considered together, one should correct for multiple comparisons using family-wise error rates (e.g. Bonferroni correction) or false discovery rates.

\section{Estimation techniques}
\label{app:estimationTechniques}

While the mathematical formulation of the quantities in \appRefSec{measures} are relatively straightforward, empirically estimating them in practice from a finite number $N$ of samples of time-series data can be a complex process, and is dependent on the type of data you have and its properties.
Estimators are typically subject to bias and variance due to finite sample size.

In this section, we introduce the various types of estimators which are included in JIDT.
Such estimators are discussed in some depth by \citet{vic14a}, for the transfer entropy in particular.
Unless otherwise noted, all quoted features and time-complexities are as implemented in JIDT.

\subsection{Discrete-valued variables}
\label{app:discreteEstimator}

For discrete variables $X$, $Y$, $Z$ etc, the definitions in \appRefSec{measures} may be used directly, by counting the matching configurations in the available data to obtain the relevant plug-in probability estimates (e.g. $\hat{p}(x \mid y)$ and $\hat{p}(x)$ for MI).
This approach may be taken for both local and average measures.
These estimators are simple and fast, being implemented in $\operatorname{O}\left(N \right)$ time even for measures such as transfer entropy which require embedded past vectors (since these may be cached and updated at each step in a time-series).
Several bias correction techniques are available, e.g. \citep{pan03,bon08a}, though not yet implemented in JIDT.

\subsection{Continuous-valued variables}
\label{app:continuousEstimators}

For continuous variables $X$, $Y$, $Z$, one could simply discretise or bin the data and apply the discrete estimators above.
This is a simple and fast approach ($\operatorname{O}\left(N \right)$ as above), though it is likely to sacrifice accuracy.
Alternatively, we can use an estimator that harnesses the continuous nature of the variables, dealing with the differential entropy and probability density functions.
The latter is more complicated but yields a more accurate result.
We discuss several such estimators in the following.
Note that except where otherwise noted, JIDT implements the most efficient available algorithm for each estimator.

\subsubsection{Gaussian-distribution model}
\label{app:gaussianEstimators}

The simplest estimator uses a \emph{multivariate Gaussian model} for the relevant variables, assuming linear interactions between them.
Under this model, for $\mathbf{X}$ (of $d$ dimensions) the entropy has the form \citep{cover91}:
\begin{align}
	\entropy{\mathbf{X}} = \frac{1}{2} \ln{ ((2 \pi e)^d \mid \Omega_\mathbf{X} \mid)}
	\label{eq:gaussianEntropy},
\end{align}
(in \textit{nats}) where $\mid \Omega_\mathbf{X} \mid$ is the determinant of the $d \times d$ covariance matrix $\Omega_\mathbf{X} = \overline{\mathbf{X} \mathbf{X}^T}$, and the overbar ``represents an average over the statistical ensemble'' \citep{barn09b}.
Any standard information-theoretic measure in \appRefSec{measures} can then be obtained from sums and differences of these joint entropies.
For example, \citet{kais02} demonstrated how to compute transfer entropy in this fashion.
These estimators are fast ($\operatorname{O}\left(N d^2 \right)$) and parameter-free, but subject to the linear-model assumption.

Since PDFs were effectively bypassed in \eq{gaussianEntropy}, the local entropies (and by sums and differences, other local measures) can be obtained by first reconstructing the probability of a given observation $\mathbf{x}$ in a multivariate process with covariance matrix $\Omega_\mathbf{X}$:
\begin{align}
	p(\mathbf{x}) = \frac{1}{(\sqrt{2 \pi})^d \mid \Omega_\mathbf{X} \mid^{1/2}} \exp{\left( -\frac{1}{2} (\mathbf{x} - \mathbf{\mu})^T \Omega_\mathbf{X}^{-1} (\mathbf{x} - \mathbf{\mu}) \right)}
	\label{eq:gaussianPDFs},
\end{align}
(where $\mathbf{\mu}$ is the vector of expectation values of $\mathbf{x}$), then using these values directly in the equation for the given local quantity as a plug-in estimate \citep{liz14b}.\footnote{This method can produce a local or pointwise Granger causality, as a local transfer entropy using a Gaussian model estimator.}

\subsubsection{Kernel estimation}
\label{app:boxKernelEstimators}

Using \emph{kernel-estimators} (e.g. see \citet{schr00} and \citet{kantz97}), the relevant joint PDFs (e.g. $\hat{p}(x, y)$ and $\hat{p}(x)$ for MI) are estimated with a \emph{kernel function} $\Theta$, which measures ``similarity'' between pairs of samples $\{ x_n,y_n\}$ and $\{ x_{n'},y_{n'}\}$ using a resolution or \textit{kernel width} $r$.
For example, we can estimate:
\begin{eqnarray}
		\hat{p}_r(x_n, y_n) = 
		 \frac{1}{N} \sum_{n'=1}^{N}{
					\Theta \left( \left| \left( 
										\begin{array}{cc}
										  x_{n} - x_{n'} \\
											y_{n} - y_{n'}
										\end{array}
					 			\right) \right| - r \right) }
	\label{eq:kernelEstimation}.
\end{eqnarray}
By default $\Theta$ is the step kernel ($\Theta(x > 0) = 0$, $\Theta(x \leq 0) = 1$), and the norm $\left| \: \cdot \: \right|$ is the maximum distance.
This combination -- a \textit{box kernel} -- is what is implemented in JIDT.
It results in $\hat{p}_r(x_n, y_n)$ being the proportion of the $N$ values which fall within $r$ of $\{x_n, y_n\}$ in both dimensions $X$ and $Y$.
Different resolutions $r$ may be used for the different variables, whilst if using the same $r$ then prior normalisation of the variables is sensible.
Other choices for the kernel $\Theta$ and the norm $\left| \: \cdot \: \right|$ are possible.
Conditional probabilities may be defined in terms of their component joint probabilities.
These plug-in estimates for the PDFs are then used directly in evaluating a local measure for each sample $n \in \left[ 1, N \right]$ and averaging these over all samples, i.e. via \eq{localEntropyExpectationN} for $\entropy{X}$ rather than via \eq{entropy} (e.g. see \citet{kais02} for transfer entropy).
Note that methods for bias-correction here are available for individual entropy estimates (e.g. as proposed by \citet{grass88} for the box kernel), but when combined for sums of entropies (as in MI, TE, etc.) \citet{kais02} state: ``this approach is not viable \ldots since the finite sample fluctuations \ldots are not independent and we cannot correct their bias separately''.\footnote{As such, these are not implemented in JIDT, except for one method available for testing with the kernel estimator for TE.} Such issues are addressed by the Kraskov-St\"{o}gbauer-Grassberger estimator in the next section.

Kernel estimation can measure non-linear relationships and is model-free (unlike Gaussian estimators), though is sensitive to the parameter choice for resolution $r$ \citep{schr00,kais02} (see below), is biased and is less time-efficient.
Naive algorithms require $\operatorname{O}\left(N^2 \right)$ time, although efficient neighbour searching can reduce this to $\operatorname{O}\left(N \log{N} \right)$ or via box-assisted methods to $\operatorname{O}\left(N \right)$ \citep{kantz97}.\footnote{These quoted time complexities ignore the dependency on dimension $d$ of the data, but will require a multiplier of at least $d$ to determine norms, with larger multipliers perhaps required for more complicated box-assisted algorithms.}
Box-assisted methods are used in JIDT for maximal efficiency.

Selecting a value for $r$ can be difficult, with a too small value yielding undersampling effects (e.g. MI the values diverge \citep{schr00}) whilst a too large value ignores subtleties in the data.
One can heuristically determine a lower bound for $r$ to avoid undersampling.
Assuming all data are normalised (such that $r$ then refers to a number of standard deviations) and spread somewhat evenly, the values for each variable roughly span 6 standard deviations and a given sample has $\sim N/(6/2r)$ coincident samples in any given dimension or $\sim N/(6/2r)^d$ in the full joint space of $d$ dimensions.
Requiring some number $K$ of coincident samples on average within $r$ (\citet{lung05} suggest $K \geq 3$ though at least 10 is more common), we then solve for $K \leq N/(6/2r)^d$.\footnote{More formally, one can consider the average number of coincidences for the \emph{typical set}, see \citet{cover91} and \citet{mart94}.}
Even within these extremes however, the choice of $r$ can have a very large influence on the comparative results of the measure; see \citet{schr00} and \citet{kais02}.

\subsubsection{Kraskov-St\"{o}gbauer-Grassberger (KSG) technique}
\label{app:ksgEstimators}

\citet*{kra04a} (KSG) (see also \citet{kra04b}) improved on (box) kernel estimation for MI by combining several specific enhancements designed to reduce errors when handling a small number of observations.
These include: the use of Kozachenko-Leonenko estimators \citep{koza87} of log-probabilities via nearest-neighbour counting; bias correction;
\todo{Should bias correction be cited to Grassberger somehow?}
and a fixed number $K$ of nearest-neighbours in the full $X$-$Y$ joint space.
The latter effectively means using a dynamically altered kernel width $r$ to adjust to the density of samples in the vicinity of any given observation, which smooths out errors in the PDF estimation.
For each sample $\{ x,y\}$, one finds the $K$th nearest neighbour in the full $\{ x,y\}$ space (using max norms to compare $x$ and $y$ distances), and sets kernel widths $r_x$ and $r_y$ from it.
The authors then propose two different algorithms for determining $r_x$ and $r_y$ from the $K$th nearest neighbour.

For the first KSG algorithm, $r_x$ and $r_y$ are set to the maximum of the $x$ and $y$ distances to the $K$th nearest neighbour, and one then counts the number of neighbours $n_x$ and $n_y$ strictly within these widths in each marginal space.
Then the averages of $n_x$ and $n_y$ over all samples are used to compute:
\begin{align}
	\mutualInfoUpper{(1)}{X}{Y} = \psi(K) - \left\langle \psi(n_x + 1) + \psi(n_y + 1) \right\rangle + \psi(N)
	\label{eq:kraskov1},
\end{align}
(in \textit{nats}) where $\psi$ denotes the digamma function.

For the second KSG algorithm, $r_x$ and $r_y$ are set separately to the $x$ and $y$ distances to the $K$th nearest neighbour, and one then counts the number of neighbours $n_x$ and $n_y$ within and on these widths in each marginal space.
Again one uses the averages of $n_x$ and $n_y$ over all samples to compute (in \textit{nats}):
\begin{align}
	\mutualInfoUpper{(2)}{X}{Y} = \psi(K) - \frac{1}{K} - \left\langle \psi(n_x) + \psi(n_y) \right\rangle + \psi(N)
	\label{eq:kraskov2}.
\end{align}

Crucially, the estimator is bias corrected, and is demonstrated to be quite robust to variations in $K$ (from $K=4$ upwards, as variance in the estimate decreases with $K$) \citep{kra04a}.
Of the two algorithms: algorithm 1 (\eq{kraskov1}) is more accurate for smaller numbers of samples but is more biased, while algorithm 2 (\eq{kraskov2}) is more accurate for very large sample sizes.

The KSG estimator is directly extendible to multi-information also; see \citet{kra04b}.

Furthermore, \citet{kra04b} originally proposed that TE could be computed as the difference between two MIs (with each estimated using the aforementioned technique).
However, the KSG estimation technique has since been properly extended to conditional MI by \citet{fre07} and transfer entropy (originally by \citet{gom10a} and later for algorithm 2 by \citet{wib14c}) with single estimators.
Here for $\condMutualInfo{X}{Y}{Z}$, for each sample $\{ x,y,z \}$, one finds the $K$th nearest neighbour in the full $\{ x,y,z \}$ space (using max norms to compare $x$, $y$ and $z$ distances), and sets kernel widths $r_x$, $r_y$ and $r_z$ from it.
Following KSG algorithm 1, $r_z$ and $\{r_{xz},r_{yz}\}$ are set to the maximum of the marginal distances to the $K$th nearest neighbour, and one then counts  $\{n_z,n_{xz},n_{yz}\}$ strictly within this width (where $n_{xz}$ and $n_{yz}$ refer to counts in the joint $\{x,z\}$ and $\{ y,z \}$ joint spaces) to obtain \citep{fre07,gom10a}:
\begin{align}
	\conditionalMutualInfoUpper{(1)}{X}{Y}{Z} = \psi(K) + \left\langle \psi(n_z + 1) - \psi(n_{xz} + 1) - \psi(n_{yz} + 1)\right\rangle
	\label{eq:kraskovCondMI1}.
\end{align}
While following KSG algorithm 2, $\{r_x,r_y,r_z\}$ are set separately to the marginal distances to the $K$th nearest neighbour, and one then counts  $\{n_z,n_{xz},n_{yz}\}$ within or on these widths to obtain \citep{wib14c}:
\begin{align}
	\conditionalMutualInfoUpper{(2)}{X}{Y}{Z} = \psi(K) - \frac{2}{K} + \left\langle \psi(n_z) - \psi(n_{xz}) + \frac{1}{n_{xz}} - \psi(n_{yz})  + \frac{1}{n_{yz}}\right\rangle
	\label{eq:kraskovCondMI2}.
\end{align}

Local values for these estimators can be extracted by unrolling the expectation values and computing the nearest neighbour counts only at the given observation $\{x,y\}$, e.g. for KSG algorithm 1 \citep{liz14b}:
\begin{align}
	\localMutualInfoUpper{(1)}{x}{y} & = \psi(K) - \psi(n_x + 1) - \psi(n_y + 1) + \psi(N)
	\label{eq:kraskov1Local}, \\
	\localCondMutualInfoUpper{(1)}{x}{y}{z} & = \psi(K) + \psi(n_z + 1) - \psi(n_{xz} + 1) - \psi(n_{yz} + 1)
	\label{eq:kraskov1LocalCond}.
\end{align}
This approach has been used to estimate local transfer entropy by \citet{liz11f} and \citet{steeg13a}.

KSG estimation builds on the non-linear and model-free capabilities of kernel estimation to add bias correction, better data efficiency and accuracy, and being effectively parameter-free (being relatively stable to choice of $K$).
As such, it is widely-used as best of breed solution for MI, conditional MI and TE for continuous data; see e.g. \citet{wib14c} and \citet{vic14a}.
On the downside, it can be computationally expensive with naive algorithms requiring $\operatorname{O}\left(K N^2 \right)$ time (again ignoring the dimensionality of the data) though fast nearest neighbour search techniques can reduce this to $\operatorname{O}\left(K N \log{N} \right)$.
For release v1.0 JIDT only implements a naive algorithm, though fast nearest neighbour search is implemented and available via the project SVN, and as such will be included in future releases.

\todo{Include diagram from my ppts? Probably won't bother}

\subsubsection{Permutation entropy and symbolic TE}
\label{app:permutationEstimators}

\emph{Permutation entropy} approaches \citep{bandt02} estimate the relevant PDFs based on the relative ordinal structure of the joint vectors (this is not suitable for PDFs of single dimensional variables).
That is, for a joint variable $\mathbf{X}$ of $d$ dimensions, a sample $\mathbf{x}$ with components $x_i$ ($i \in \{ 0 \ldots d-1 \}$) is replaced by an ordinal vector $\mathbf{o}$ with components $o_i \in \{ 0 \ldots d-1 \}$, where the value of $o_i=r$ assigned for $x_i$ being the $r$-th largest component in $\mathbf{x}$.
The PDF $p(\mathbf{x})$ is replaced by computation of $\hat{p}(\mathbf{o})$ for the corresponding ordinal vector, and these are used as plug-in estimates for the relevant expected or local information-theoretic measure.

Permutation entropy has for example been adapted to estimate TE as the \textit{symbolic transfer entropy} \citep{stan08a}, with local symbolic transfer entropy also defined \citep{nak12a,nak13a}.

Permutation approaches are computationally fast, since they effectively compute a discrete entropy after the ordinal symbolisation ($\operatorname{O}\left(N \right)$).
They are a model-based approach however, assuming that all relevant information is in the ordinal relationship between the variables.
This is not necessarily the case, and can lead to misleading results, as demonstrated by \citet{wib13a}.